\documentclass[aps,prd,preprintnumbers,groupedaddress,nofootinbib,amssymb,eqsecnum,notitlepage,10pt]{revtex4-2}

\usepackage{graphicx}
\usepackage{color} 
\usepackage{dcolumn}
\usepackage{amsmath,amsthm,amssymb}
\usepackage{amsfonts}
\usepackage{bm}
\usepackage{mathrsfs}
\usepackage{revsymb}
\usepackage{comment}
\usepackage[bookmarks=true,bookmarksnumbered=true,colorlinks=false]{hyperref}
\newif\ifoneauthor
\oneauthortrue

\begin{document}

\title{Strong lensing of gravitational waves with modified propagation}

\author{Hiroki Takeda}
\email[]{takeda@tap.scphys.kyoto-u.ac.jp}
\affiliation{The Hakubi Center for Advanced Research, Kyoto University, Kyoto 606-8501, Japan}
\affiliation{Department of Physics, Kyoto University, Kyoto 606-8502, Japan}
\author{Takahiro Tanaka}
\affiliation{Department of Physics, Kyoto University, Kyoto 606-8502, Japan}
\affiliation{Center for Gravitational Physics and Quantum Information, Yukawa Institute for Theoretical Physics, Kyoto University, Kyoto 606-8502, Japan}

\date{\today}

\begin{abstract}
\noindent
We explore the impact of corrections to the propagation on the waveforms of gravitationally lensed gravitational waves under the geometrical optics approximation, focusing on both uniform cosmological modifications and local modifications localized around lensing objects. By adopting a model-independent phenomenological approach,
we systematically investigate the effects of these modifications in strong lensing scenarios, where detection of multiple images is expected. Our analysis reveals that cosmological modifications can yield corrections to the time delay that remain to be minor compared with the effects that accumulate over the whole propagation process, which are present also in the unlensed waveform. By contrast, local modifications around lensing objects can alter the image position and also the magnification factor, which is potentially polarization-selective and frequency-dependent. In some case we can have image disappearance as well as signal amplification. Furthermore, we demonstrate that such modifications can cause degradation of waveform match with the templates based on general relativity.
This study highlights the importance of considering waveform modifications to search for the signature of modified propagation or the existence of extra polarization modes, and proposes potential observational targets.
\end{abstract}

\maketitle


\section{Introduction}
Gravitational lensing is a phenomenon in which the path and intensity characteristics of propagating waves are altered by the gravitational field of intervening objects~\cite{Schneider:1992bmb, dodelson:2017, Bartelmann:2010fz}. 
In the case of electromagnetic waves, gravitational lensing can lead to phenomena such as magnification, multiple images, or arc-shaped images known as Einstein rings. Gravitational lensing of electromagnetic waves has enabled the observation of distant or dim celestial bodies, the measurement of mass distribution through weak lensing effects, and the search for exoplanets using microlensing techniques. 

Similarly, gravitational waves (GWs) are expected to be affected by gravitational lensing~\cite{Ohanian:1974ys}. The impact of gravitational lensing on GWs depends on the properties of the lensing object. For sufficiently massive lenses, GWs can experience amplification without altering the waveform, potentially resulting in multiple signals with time delays, ranging from minutes to months for galaxy lenses, and up to several years for galaxy cluster lenses~\cite{Wang:1996as, Dai:2017huk, Ezquiaga:2020gdt}. In the case of low mass lenses, microlensing can cause interference patterns in the waveforms~\cite{Nakamura:1997sw, Takahashi:2003ix, Cao:2014oaa, Jung:2017flg, Lai:2018rto, Christian:2018vsi, Diego:2019lcd, Cheung:2020okf}.

The LIGO-Virgo detectors~\cite{LIGOScientific:2014pky, VIRGO:2014yos} have reported 83 binary black hole merger, 2 binary neutron star merger, and 5 neutron star-black hole merger event candidates in the run event catalog, GWTC3~\cite{KAGRA:2021vkt}. Currently, the fourth observing run, O4, with the participation of the KAGRA detector~\cite{KAGRA:2020tym}, is underway. This time 81 compact binary coalescence events have already been alerted. Assuming that binary black hole merger rate reflects the star formation rate density, the detection of strongly lensed GWs is expected within the operation period of LIGO-Virgo-KAGRA network~\cite{Ng:2017yiu, Li:2018prc, Oguri:2018muv}. Analyses have been conducted using data collected so far to search for signatures of gravitational lensing, but clear evidence has yet to be explored~\cite{LIGOScientific:2021izm, LIGOScientific:2023bwz}.

The detection of lensed GWs can be expected to contribute to the tests of general relativity (GR). For instance, tests of the propagation speed of GWs relative to electromagnetic waves~\cite{Baker:2016reh, Collett:2016dey, Fan:2016swi}, separation of polarizations~\cite{Goyal:2020bkm}, and birefringence effects~\cite{Goyal:2023uvm} have been reported. 
Gravitational lensing phenomena have been investigated beyond GR~\cite{Ezquiaga:2020dao, Streibert:2024cuf}.
Generally, in theories beyond GR, the propagation equations of GWs are modified~\cite{Hwang:1996xh, Will:1997bb, Mukohyama:2009zs, Sefiedgar:2010we, Vacaru:2012nvq, Saltas:2014dha, Yunes:2016jcc, Kostelecky:2016kfm, Max:2017flc, Nishizawa:2017nef}. Tests assuming free propagation with cosmological long-distance modifications have been conducted for GWTC3 events~\cite{LIGOScientific:2021sio}. 
Such modifications are expected to also affect lensed GWs.
Besides this possibility, extensions of gravity theory may provide modifications to GW propagation localized around lensing objects. 
For example, the effective field theory of gravity predicts a change of the GW speed around a black hole~\cite{deRham:2020ejn}.
Scalar-tensor theories predict screening mechanisms around massive objects that shield the fifth force, thereby restoring GR predictions~\cite{Khoury:2003aq, Khoury:2003rn, Vainshtein:1972sx, Deffayet:2001uk, Hinterbichler:2010es, Hinterbichler:2011ca, Babichev:2009ee, Babichev:2013usa}. Although the mechanism of the screening depends on the underlying theory, typically it is associated with non-trivial background configuration that is likely to affect the propagation of scalar modes~\cite{Ezquiaga:2020dao}.
In massive gravity, graviton mass can be much larger in the strong gravity regime such as in the vicinity of black holes~\cite{Zhang:2017jze}.
Such local modifications are also expected to contribute to the gravitational lensing of GWs. Hence, one may be able to extract information about the exotic properties around massive objects from observations of lensed GWs. 
Specifically, these modifications should become manifest as changes in the deflection potential, which may lead to a variety of effects on the waveform.

In this paper, we investigate phenomenologically the effects of the modified propagation on the waveform of lensed GWs in the strong lensing regime, in which multiple images are isolated in time from each other in the geometrical optics approximation. 
We systematically formulate how the waveform of GWs is affected by strong gravitational lensing for future tests of GR. 
In particular, we are interested in the possibility that modified propagation amplifies certain polarization modes of GWs selectively, which motivates the search for non-tensorial modes beyond GR~\cite{LIGOScientific:2018czr,Hagihara:2019ihn,Takeda:2020tjj,Takeda:2021hgo,LIGOScientific:2021sio}. 
The amplitude of non-tensorial modes should be suppressed in the unlensed GW signal to be consistent with the existing experiments~\cite{Takeda:2023mhl}.
While cosmological modifications predominantly alter the time delay, local modifications can alter the image position and the magnification factor. 
At the same time, the local modifications introduce new frequency dependence.
We will find that the modified waveforms largely deviate from the GR ones, and hence dedicated new set of templates are required for the search of lensed GWs.

In Sec.~\ref{sec:propagation_eq}, we present the effective propagation equation for GWs propagating on the background around a lensing object. 
Sec.~\ref{sec:amplification_factor} provides expressions for the transfer function, known as the amplification factor, for GW waveforms lensed with spherically symmetric local modifications as well as those with cosmological modifications. 
In Sec.~\ref{sec:time_delay_and_magnification_factor}, we assume a specific lens and modified propagation models and calculate the time delay and magnification factor. 
Sec.~\ref{sec:detectability} discusses detectability by evaluating the signal-to-noise ratio and the match between the GR and modified waveforms.
Finally, we devote Sec.~\ref{sec:discussion} and Sec.~\ref{sec:conclusion}, respectively, to the discussion and conclusion of the paper.
In this paper, we use the units in which $c=G=1$. 
For any function of the spacetime coordinates $(t, x^i)$, we define the temporal Fourier transformation as
\begin{align}
    \tilde{f}(\omega,x^i):=\frac{1}{2\pi}\int f(t, x^i)e^{+i\omega t} dt\,.
\end{align}

\section{Effective propagation equation of gravitational waves}
\label{sec:propagation_eq}
Let us consider canonically normalized fields $\phi_I$ corresponding to $N$ independent propagating polarization modes of gravitational waves. We assume that the metric perturbation $h_{\mu\nu}$ is written in terms of $\phi_I$ under the synchronous gauge $h_{0\mu}=0$ as
\begin{align}
    h_{ij}=\phi_{I}E^{I}_{ij}\,,
    \label{eq:propagating_modes}
\end{align}
where $E^{I}_{ij}$ are polarization tensors for $\phi_I$. 
The polarization tensors $E^{I}_{ij}$ can be written as 
a linear combination of the standard polarization basis tensor $e^{A}_{ij}$, 
\begin{align}
    E^{I}_{ij}=\xi^{I}_{A}(x) e^{A}_{ij}\,. 
    \label{eq:polarization_tensors}
\end{align}
The expansion coefficients $\xi^{I}_{A}(x)$ are referred to as the coupling parameters. 
The standard polarization basis tensors $e^{A}_{ij}$ are the symmetric tensors composed of irreducible tensor representations of SO(2) around the propagation direction of the GW specified by the unit vector in the propagation direction $\hat{k}^{i}$.
The signal from GW detectors can be expressed as a summation of the GW polarizations weighted by the corresponding detector's response functions.
Please refer to Appendix~\ref{sec:polarization_and_signal} for the details of GW polarizations and GW signal of interferometric detectors.
In this paper, we will discuss the evolution of each mode affected by the modifications to the propagation around lensing objects.

\begin{figure}
    \centering
    \includegraphics[width=5.8in]{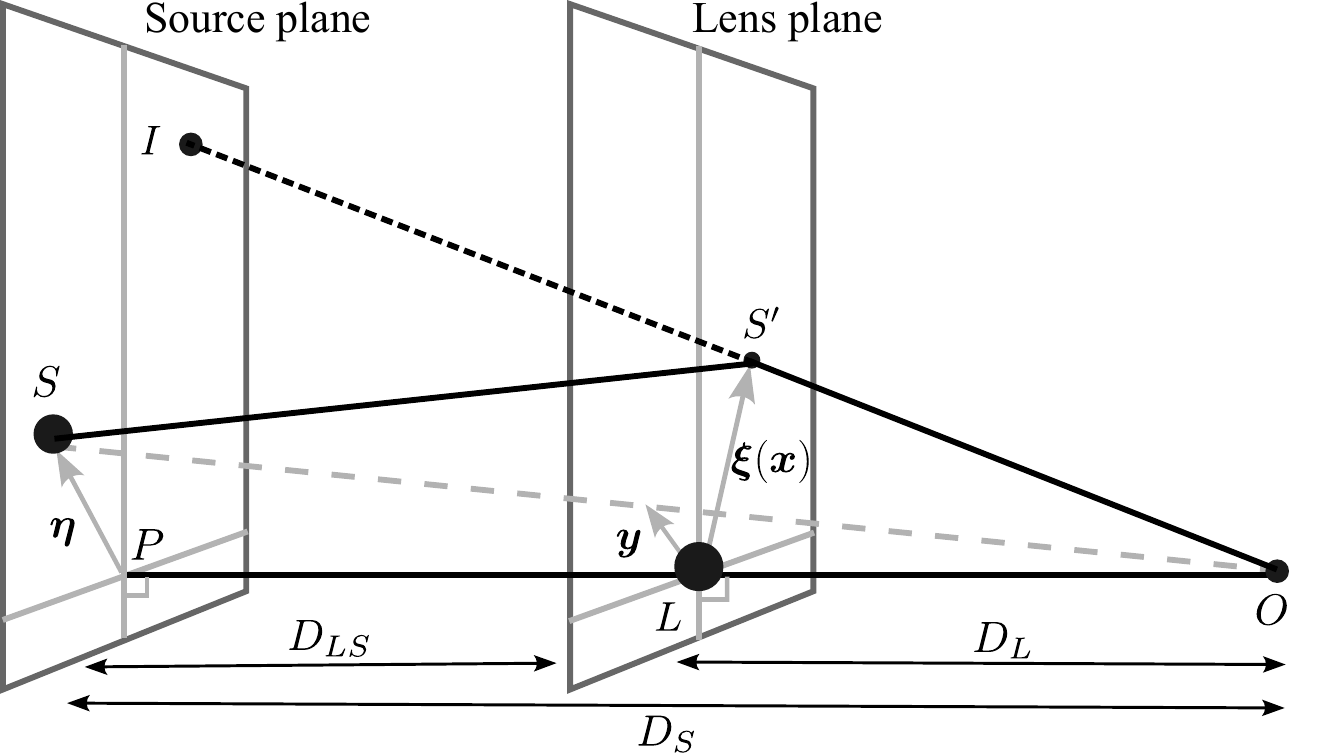}
    \caption{Our geometrical configuration of a gravitational-lensing system, composed of a GW source ($S$), a lens object ($L$), and an observer ($O$). The lens plane including $L$ and the source plane including $S$ are perpendicular to the line $OL$. $D_{L}$ is the distance between $OL$, $D_{LS}$ is the distance between the lens plane and the source plane, and $D_S=D_L + D_{LS}$ is the source distance. The point $I$ is the position of the image induced by the gravitational lensing and $S'$ is the point of the source on the lens plane. We write the position vector of $S'$ on the lens plane as $\bm{\xi}$ and of $S$ on the source plane as $\bm{\eta}$. The position vectors normalized and projected onto the lens plane are denoted as $\bm{x}$ and $\bm{y}$.}
    \label{fig:sketch}
\end{figure}

We consider a gravitational-lensing system as described in the Fig~\ref{fig:sketch}. We have a GW source at a distance $D_S$ from the observer $O$, and a lens object $L$ between the source and the observer, at a distance $D_L$ from the observer. $D_{LS}$ is the distance between the source and the lens. 
Here, $D_{S}$, $D_{L}$, and $D_{LS}$ are the angular diameter distances. The plane including $S$ perpendicular to the line of sight is called the source plane and the plane including $L$ the lens plane. Denoting the point at which the line connecting $O$ and $L$ intersects the source plane as $P$, we refer to the vector from $P$ to $S$ as $\bm{\eta}$ and the position vector relative to $L$ in the lens plane as $\bm{\xi}$. We introduce 
\begin{align}
    \bm{x}:= \frac{\bm{\xi}}{\xi_0}, \qquad \bm{y}:=\frac{D_L}{\xi_0 D_S}\bm{\eta}\,, 
\end{align}
by normalizing $\bm{\xi}$ and $\bm{\eta}$ using an arbitrary reference distance scale $\xi_0$ and projecting them onto the lens plane. 

On general backgrounds, different polarization modes can couple with 
each other.
In the present treatment the complications coming from the mode mixing are encapsulated in $\xi^I_A(x)$, which can be frequency dependent~\cite{Takeda:2023mhl}.
We phenomenologically consider the following effective propagation equation for one propagating mode $\phi$ in Fourier space,
\begin{align}
    \left[\omega^2 +\partial_i \partial^i  - M^2(\omega, x^i) \right]\tilde{\phi}(\omega, x^i) = 4\omega^2 U \left(1- \frac{N^2(\omega, x^i)}{\omega^2} \right)\tilde{\phi}(\omega, x^i) \,.
\label{eq:propagation_eq}
\end{align}
where $M^2$ and $N^2$ represent the effects of gravity modifications. 
For example, they take the form of 
\begin{align}
M^2(\omega, x^i)=c_{\phi}^2(x^i) \omega^2 + \mu^2(x^i)\,,\quad
{N}^2(\omega, x^i)=c_{\phi}^2(x^i) \omega^2 + \frac{\mu^2(x^i)}{2}\,.  
\end{align}
Here, $c_{\phi}(x^i)$ represents the deviation of the propagation speed from the speed of light and $\mu(x^i)$ gives the mass of the $\phi$ mode. 
In general, we can consider more general frequency dependence for 
$M^2$ and $N^2$. 

$N^2$ is not independent of but not identical to $M^2$, and the relation between them depends on the type of interaction. 
However, as will become apparent, the effects of $N^2$ are always subdominant, making their specific form unimportant.
We define the phase velocity by
\begin{align}
    v_{p} = 1 + \frac{M^2}{2\omega^2} + 2U-\frac{2UN^2}{\omega^2}\,.
\end{align}
We assume that the parameters such as $c_\phi^2$ and $\mu^2$ are either almost spatially homogeneous, or spherically symmetric and localized around lens objects.
The results in GR are recovered in the limit of setting the modification parameters $M(\omega, x^i)$ and $N(\omega, x^i)$ to zero.

\section{Amplification factor with modified propagation equation}
\label{sec:amplification_factor}
To predict the observed waveforms in different situations, we need to solve the differential equations,
\begin{align}
    \left[\omega^2 +\partial_i \partial^i - M^2_{\rm eff}(\omega)\right]\tilde{\phi}(\omega, x^i) = 4\omega^2 {U}_{\rm eff}(\omega, x^i) \tilde{\phi}(\omega, x^i) \,,
\label{eq:solve_propagation_eq}
\end{align}
where the effective mass $M_{\rm eff}(\omega)$ describes the spatially homogeneous modifications, while the effective potential ${U}_{\rm eff}(\omega, x^i)$ does the localized modifications as well as the grvitational lensing potential.  
Adopting thin lens approximation, we assume that $U_{\rm eff}$ is at work only around the lens plane. 
Hence, away from the lens plane, the effective phase velocity is reduced to
\begin{align}
    v_{p, {\rm eff}} = 1 + \frac{M^2_{\rm eff}}{2\omega^2}\,. 
\end{align}

Applying Green's theorem to the volume surrounding the observer for $\tilde{\phi}$ and the Green's function $e^{i(\omega/v_{p, {\rm eff}}) |{\bm x}-{\bm x}_S|}/|{\bm x}-{\bm x}_S|$ to the homogeneous equation of Eq.~\eqref{eq:solve_propagation_eq}, we find the Fresnel-Kirchhoff diffraction integral over the lens plane. 
Finally, we obtain the transfer function, so-called the amplification factor,
\begin{align}
    F(f)&:=\tilde{\phi}^L(f)/\tilde{\phi}^{UL}(f)\, \nonumber \\
    &=\xi^{2}_{0}\frac{D_{S}}{D_{L}D_{LS}}\frac{f}{i v_{p, {\rm eff}}
    }\int d^2x \exp[2\pi i f T(\bm{x}, \bm{y})]\,,
    \label{eq:transfer_function}
\end{align}
which is defined by the ratio of the lensed waveform $\tilde{\phi}^{L}$ to the unlensed one $\tilde{\phi}^{\rm UL}$, 
where $f=\omega/(2\pi)$ is the GW frequency and $T(\bm{x}, \bm{y})$ is the Fermat potential, {\it i.e.,} the time delay along the path from a normalized position $\bm y$ on the source plane to the observer via a point $\bm x$ on the lens plane. 
The time delay consists of two parts,
\begin{align}
    T(\bm{x}, \bm{y}) = \Delta T_{\rm geo}(\bm{x}, \bm{y}) + \Delta T_{\rm potential}(\bm x)\,.
\end{align}
The first part
\begin{align}
    \Delta T_{\rm geo}(\bm{x}, \bm{y}) = \frac{\xi_0^2}{v_{p, {\rm eff}}} \frac{D_{S}}{D_{L} D_{LS}}\frac{1}{2}|\bm{x} - \bm{y}|^2
    \label{eq:geometrical_time_delay}
\end{align}
is the geometrical time delay 
evaluated by assuming a flat spacetime. 
The other is the time delay due to the potential,
\begin{align}
    \Delta T_{\rm potential}(\bm x) = -2 \int U_{\rm eff}(\bm{x}, \ell) d\ell\,,
    \label{eq:potential_time_delay}
\end{align}
where the integration is performed along the propagation path. 
Following the standard convention, we express the time delay as 
\begin{align}
T(\bm{x}, \bm{y})= \frac{\xi_0^2}{v_{p, {\rm eff}}}\frac{D_{S}}{D_{L} D_{LS}}\left[\frac{1}{2}|\bm{x} - \bm{y}|^2 - \psi(\bm{x})\right]\,,
\end{align}
by introducing the deflection potential 
\begin{align}
    \psi(\bm{x}):= -\frac{v_{p, {\rm eff}}}{\xi_0^2} \frac{D_L D_{LS}}{D_S} \Delta T_{\rm potential}
      =\frac{2 v_{p, {\rm eff}}}{\xi_0^2} \frac{D_L D_{LS}}{D_S}  \int U_{\rm eff}(\bm{x}, \ell) d\ell\,.
\end{align}

In the geometrical optics limit, in which the frequency is high enough compared to the time delay, the contributions from the stationary points in the integration in Eq.~\eqref{eq:transfer_function} dominate. 
The stationary points, which we call the lens image positions, satisfy 
\begin{align}
    \left.\frac{\partial T({\bm{x}, \bm{y}})}{\partial \bm{x}}\right|_{\bm{x}=\bm{x}_j}=0\,,
\end{align}
which gives the lens equation
\begin{align}
    \bm{x}_{j}-\bm{y}=\frac{\partial \psi}{\partial {\bm x}}(\bm{x}_{j}, \bm{y})\,.
\end{align}
In the stationary phase approximation we expand $T(\bm{x}, \bm{y})$ around each image position 
to the quadratic order, 
and perform the Gaussian integrals, to obtain 
\begin{align}
    F(f)&=  |\mu_{\rm mag}(\bm{x}_j)|^{1/2}\exp[2\pi i f T(\bm{x}_j, \bm{y}) - i\pi n_j]\,,
\end{align}
where 
\begin{align}
    \mu_{\rm mag}(\bm{x}_j)= \frac{1}{|\lambda_{1}(\bm{x}_1)\lambda_{1}(\bm{x}_2)|}\,
\end{align}
is the magnification factor,
with $\lambda_{1}(\bm{x}_j)$ and $\lambda_{2}(\bm{x}_j)$ being the eigenvalues of the Hessian matrix of $T(\bm{x}, \bm{y})$ 
evaluated at $\bm{x}_j$. 
$\exp[-i\pi n_j]$ is the Morse phase, and $n_j$ takes the value of $0$ (minimum point), $1/2$ (saddle point), or $1$ (maximum point), depending on the combination of the signs of the eigenvalues of the Hessian matrix. 

Below, we introduce two types of models of modified GW propagation: (i) cosmological uniform modifications and (ii) spherically symmetric local modifications, and later compare their observational consequences.

\subsection{Cosmological modification}
First, we consider cosmological uniform modifications assuming that $M^2(\omega, x^i)$ is independent of $x^i$ during propagation, 
\begin{align}
    M^2(\omega) = c_{\phi}^2 \omega^2 + \mu^2\,,
    \label{eq:cosmological_modification}
\end{align} 
where $c_{\phi}$ and $\mu$ are constant parameters. 
Hence, the effective mass $M_{\rm eff}$ is identical to $M(\omega)$ and the effective phase velocity becomes
\begin{align}
    v_{p, {\rm eff}} = 1+\frac{M^2}{2\omega^2}\,,
    \label{eq:phase_velocity_cosmological}
\end{align} 
while the effective potential is the gravitational potential scaled as
\begin{align}
     U_{\rm eff}= U\left(1+\frac{N^2}{\omega^2}\right)^{-1}\,.
     \label{eq:effective_potential_cosmological}
\end{align}

The geometrical time delay is given by Eq.~\eqref{eq:geometrical_time_delay} with the phase velocity~\eqref{eq:phase_velocity_cosmological}.  
Substituting Eq.~\eqref{eq:effective_potential_cosmological} into Eq.~\eqref{eq:potential_time_delay}, we obtain the deflection potential $\psi(\bm{x})$ as
\begin{align}
    \psi_{\rm GL}({\bm x})=\frac{1}{\pi \Sigma_{\rm cr}} \left(\frac{1+M^2/\omega^2}{1+N^2/\omega^2}\right)\int d^2{\bm x}' \Sigma(\bm{x}')\left[\ln{|{\bm x}-{\bm x}'|}-\frac{1}{2}\log{\left(\frac{4D_L D_{LS}}{\xi_0^2}\right)} \right]\,,
    \label{eq:deflection_potential_GL}
\end{align}
where the surface density $\Sigma(\bm{x}')$ is defined by the integration of the the mass density $\rho$ along the direction perpendicular to the lens plane,
\begin{align}
    \Sigma({\bm x'}):= \int \rho(\xi_0 {\bm x'}, \ell') d\ell'\,,
\end{align}
and the critical density is defined by
\begin{align}
    \Sigma_{\rm cr}:=\frac{D_S}{4\pi D_L D_{LS}}\,.
\end{align}

From the above expressions, we find that, besides the overall scaling of the time delay by the constant phase velocity, the deflection potential is modified by the factor $\simeq 1+\mu^2/(2\omega^2)$.
If the modification is only in the change of phase velocity $c_{\phi}^2$, this factor becomes unity, 
and hence considering lensing systems does not add any characteristics in the waveforms.
By contrast, $\mu^2$ mass term changes the image positions, the time delays, and the magnification factors in a frequency-dependent manner.
Although this effect is unique to the lensed waveform, its observational significance would be limited. 
This limitation arises because, when such modifications introduce additional signatures to the lensed waveform, the inherent effects from the same modifications accumulate over the much longer cosmological distances in the unlensed waveform, becoming extremely pronounced. 
This can lead to conflicts with existing observations or make observations challenging. 
Consequently, the contributions of these modifications to the amplification factor would be overshadowed by the more substantial effects that are present already in unlensed waveforms.

\subsection{Spherically symmetric local modification}
\label{sec:spherical}
Next, let us consider spherically symmetric local modifications. We assume that 
$M^2(\omega, x^i)$ is spherically symmetric functions and take non-zero values only around the lens object. 
In this case, the effective phase velocity is not modified $v_{p, {\rm eff}}=1$. 
Since the effective potential is modified as
\begin{align}
    U_{\rm eff}= U + \frac{M^2}{4\omega^2}\,,
\label{eq:effective_potential_spherical}
\end{align}
the deflection potential becomes 
\begin{align}
    \psi = \psi_{GL} + \psi_{M}\,, 
\end{align}
with the additional contribution 
\begin{align}
    \psi_M=\frac{1}{\xi_0^2}\frac{D_L D_{LS}}{D_S} \int \frac{M^2}{2\omega^2} d\ell \,, 
    \label{eq:deflection_potential_spherical}
\end{align}
which can be frequency dependent.
Here, we have neglected the second-order term of $-UM^2/2\omega^2$ in Eq.~\eqref{eq:effective_potential_spherical}, assuming that both $U$ and $M^2/\omega^2$ are small.

In summary, spherically symmetric local modifications in GW propagation introduce the frequency dependent effective deflection potential. 
Consequently, image positions, time delays, and magnification factors all can vary in a frequency-dependent manner.
In contrast to the cosmological modifications, the detectability of the local modifications is not constrained yet. 

\section{Time delay and magnification for point mass model}
\label{sec:time_delay_and_magnification_factor}
In this section, we specifically compare the two different types of modifications provided in Section~\ref{sec:amplification_factor}, focusing on time delays and magnification factors. 
For demonstration, we assume a point mass lens model, for simplicity.  
The surface density is given by $\Sigma(\bm{\xi})=M_L \delta (\bm{\xi})$ where $M_L$ is the mass of the lens object. 
In the following discussion, we set the normalization length $\xi_0$ to the Einstein radius:
\begin{align}
    \xi_0=\sqrt{\frac{4M_L D_L D_{LS}}{D_S}}\,.
\end{align}

\subsection{Cosmological modification}
Under the assumption of the cosmological modifications given by Eq.~\eqref{eq:cosmological_modification}, the lens equation becomes
\begin{align}
    y=x-\left(1+\frac{\mu^2}{2\omega^2}\right)\frac{1}{x}\,, 
    \label{eq:lens_equation_cosmological}
\end{align}
where the modifications are assumed to be small. 
By solving the lens equation, we obtain the image position, 
\begin{align}
    x_{\pm, {\rm C}}=\frac{1}{2} \left(y\pm\sqrt{\frac{4\mu^2}{\omega^2} + y^2+4}\right)\,.
    \label{eq:image_position_cosmological}
\end{align}
The positive solution $x_{+, {\rm C}}$ corresponds to the image on the source side, while the negative solution $x_{-, {\rm C}}$ corresponds to the image below the optical axis.
Hence, the time delay is given by
\begin{align}
    T_{\rm C}=\frac{\xi_0^2}{v_{p, {\rm eff}}} \frac{D_{S}}{D_{L} D_{LS}} \left[\frac{1}{2} x_{\mp, {\rm C}}^2- \left(1+\frac{\mu^2}{2\omega^2}\right) \left(\log(x_{\pm, {\rm C}})-\frac{1}{2}\log{\left(\frac{4D_L D_{LS}}{\xi_0^2}\right)} \right)\right]\,.
    \label{eq:time_delay_cosmological}
\end{align}
The last term, which is just a constant, is usually omitted in GR, but it is not so in the present context because the effective phase velocity $v_{p, {\rm eff}}$ generally depends on frequency.
In this paper, 
we measure the time delay by setting the origin so that the time delay at a high frequency reference point $\sim 10^{5}\ {\rm Hz}$ vanishes.
From the eigenvalues of the Hessian matrix of the time delay, we obtain the magnification factor as,
\begin{align}
    \mu^{\rm mag}_{\pm, {\rm C}}= \left(1-\frac{(1+\mu^2/(2\omega^2))^2}{x_{\pm, {\rm GR}}^4}\right)^{-1}\,,
    \label{eq:magnification_cosmological}
\end{align}
in the geometrical optics regime.

From Eq.~\eqref{eq:time_delay_cosmological}, it is evident that the terms of $c^2_{\phi}$ and $\mu^2$ in $v_{p, {\rm eff}}$ bring the frequency dependences in the time delay, proportional to $\omega^0$ and $\omega^{-2}$, respectively. 
If we consider more general modified propagation, any kind of frequency dependence can be introduced. 
However, these effects are significantly overshadowed by the more substantial effects accumulated during cosmological long-distance propagation present in the unlensed waveforms.
The term $\mu^2$ also contributes to the second term in Eq.~\eqref{eq:time_delay_cosmological} through the effective deflection potential, and thereby affecting the magnification factor.
As discussed in Section~\ref{sec:amplification_factor}, the accumulated effect of $\mu^2$ leads to significant deformation of the unlensed waveform when this unique effect is dominant.
This can be observed from the fact that by replacing $\beta \rho_{-2, -1}$ with $-\mu^2/2$ in the equations for the image positions~\eqref{eq:image_position_cosmological}, the time delay~\eqref{eq:time_delay_cosmological}, the magnification factor~\eqref{eq:magnification_cosmological} for local modifications, we obtain the corresponding equations for the image positions~\eqref{eq:image_position_sperical}, the time delay~\eqref{eq:time_delay_spherical}, the magnification factor~\eqref{eq:magnification_factor_spherical} for cosmological modifications. 
Thus, the primary motivation for considering lensed GWs is limited to the conventional magnification of the signal, which allows us to detect events even at a higher redshift.

\subsection{Spherically symmetric local modification}
For the purpose of assessing the observational signatures of the spherically symmetric local modifications described by Eq.~\eqref{eq:effective_potential_spherical}, we specify models of $M^2(\omega,x^i)$. 
Clearly, the form of $M^2(\omega,x^i)$ depends on the theories of gravity that we concern. 
Since the aim here is not to explore a particular model, we assume here that $M$ takes a power law shape, 
with the radial and frequency power law indices $n$ and $p$ as
\begin{align}
    M^2(\omega, x^i) = \sum_{p,n} \rho_{p, n} \omega^p \left(\frac{r}{\xi_0}\right)^n\,,
\end{align}
where $\rho_{p, n}$ is the amplitude of the modification.
Note that the terms with $p=0$ and $p=-2$ correspond to $c^2_{\phi}$ and $\mu^2$, respectively.
$\rho_{p,n}$ can take any positive or negative values.

Performing the integration in Eq.~\eqref{eq:deflection_potential_spherical}, we obtain
\begin{align}
     \psi_{M}(\bm x)=\beta  \sum_{p, n} \rho_{p, n} \omega^p P_n(\bm{x})\,,
     \label{eq:effective_deflection_potential_power_law}
\end{align}
where we have introduced a constant $\beta:= (1/\xi_0) (D_L D_{LS}/ D_S)=\xi_0/(4M_L)$ and a function 
\begin{align}
    P_n(\bm{x})=S_n(\bm{x}, D_{LS})+S_n(\bm{x}, D_L)\,.
\end{align}
Here, $S_n(\bm{x}, \ell)$ is 
defined by
\begin{align}
S_n(\bm{x}, \ell) &=  \frac{1}{2}|x|^n \left(\frac{\ell}{\xi_0}\right)  {}_2 F_1\left(\frac{1}{2}, -\frac{n}{2}, \frac{3}{2}, -\frac{(\ell/\xi_0)^2}{|x|^2}\right)\,,
\label{eq:S_n}
\end{align}
where ${}_2 F_1$ is the hyper geometric function.

Phenomenologically, the simplest setup would be the case in which $M^2/4\omega^2$ is proportional to $1/r$, which results in the same form of the deflection potential as a point mass lens.
The cases with other profiles will be discussed in Sec.~\ref{sec:discussion}.
For $n=-1$, 
we have 
$S_n(\bm{x})=(1/2)\sinh^{-1}(\ell/|\bm{x}|)=(1/2)\{\log{(\ell + \sqrt{\ell^2 + |\bm{x}|^2})}-\log{|\bm{x}|}\} $, and hence under the conditions of $|\bm{x}|\ll D_{\rm L}/\xi_0$ and $|\bm{x}|\ll D_{\rm LS}/\xi_0$ the additional deflection potential is given by 
\begin{align}
    \psi_{M}(\bm x) = \beta \sigma(\omega) \left[\frac{1}{2}\log{\left(\frac{4D_L D_{LS}}{\xi_0^2}\right) - \log|\bm{x}|}\right]\,,
    \label{eq:effective_deflection_potential_n=-1}
\end{align}
with 
\begin{align}
    \sigma(\omega)=\sum_{p}\rho_{p, -1}\omega^p\,,
\end{align}
which summarizes the local modification effects. 
Accordingly, since the lens equation is modified as
\begin{align}
    y=x-\left(1-\beta \sigma(\omega)\right)\frac{1}{x}\,,
    \label{eq:lens_equation_main}
\end{align}
the image position is given by
\begin{align}
    x_{\pm, S}=\frac{1}{2} \left(y\pm\sqrt{-4 \beta\sigma(\omega) +y^2+4}\right)\,,
    \label{eq:image_position_sperical}
\end{align}
and then the time delay is written as
\begin{align}
    T_S=\xi_0^2 \frac{D_{S}}{D_{L} D_{LS}} \left[\frac{1}{2} x_{\mp, S}^2-\left(1-\beta\sigma(\omega)\right) \left( \log(x_{\pm, S})-\frac{1}{2}\log{\left(\frac{4D_L D_{LS}}{\xi_0^2}\right)} \right) \right]\,.
    \label{eq:time_delay_spherical}
\end{align}

\begin{figure}[ht]
\begin{center}
\includegraphics[width=3in]{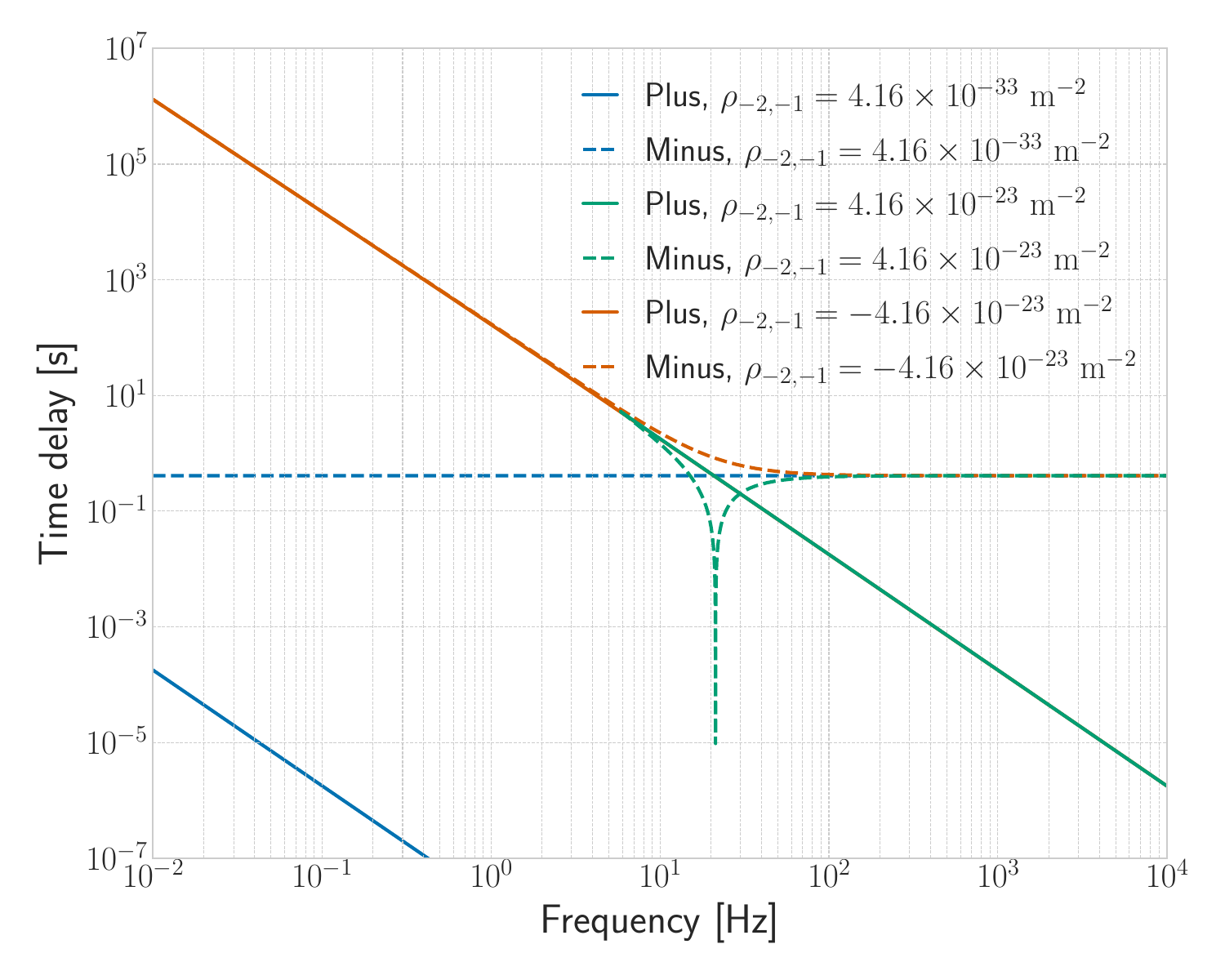}
\end{center}
\caption{Frequency dependence of time delay for non-zero $\mu$. The modification parameters are set to $\rho_{-2, -1}=4.16\times 10^{-33}\ {\rm m^{-2}}$ and $\rho_{-2, -1}=\pm4.16\times 10^{-23}\ {\rm m^{-2}}$. These correspond to masses of $\phi$ at $r=\xi_0$, $m_{\phi}=1.27\times10^{-23}\ {\rm eV}$ and $m_{\phi}=1.27\times10^{-18}\ {\rm eV}$, respectively. The fiducial values for the lensing parameters are $z_S = 0.5$, $z_L=0.1$, $M_L=10^4\ {M_{\odot}}$, and $y=1.0$.}
\label{fig:freq_vs_time_delay_spherical}
\end{figure}

Figure~\ref{fig:freq_vs_time_delay_spherical} shows the frequency dependence of the time delay for three cases. In the first case, the non-zero parameter is $\rho_{-2, -1}=4.16\times 10^{-33}\ {\rm m^{-2}}$, which is equivalent to the case with a mass of $\phi$, $m_{\phi}=1.27\times10^{-23}\ {\rm eV}$ at $r=\xi_0$. In the second and the third cases, we set $\rho_{-2, -1}=\pm 4.16\times 10^{-23}\ {\rm m^{-2}}$, corresponding to $m_{\phi}=1.27\times10^{-18}\ {\rm eV}$.
Note that $1.27 \times 10^{-23}\ \text{eV}$ is the current constraint on the graviton mass assuming simple cosmological modification~\cite{LIGOScientific:2021sio}, although this constraint does not apply to the modifications localized around massive objects even for tensorial modes as well as for non-tensorial modes.
The plotted time delay is the value after subtracting the time delay at the reference value at a high frequency for the plus image.
Unless otherwise specified, we choose the fiducial parameters of the lens object as $z_S = 0.5$, $z_L=0.1$, $M_L=10^4\ {M_{\odot}}$, and $y=1.0$. 
This choice is made to be appropriate for observing the strong gravitational lensing effects at the optimal frequencies for the second-generation GW detectors, $\sim 100\ \text{Hz}$.
Since the typical difference in the arrival times between two images is 
estimated as $\sim 4M_L = 2\times10^{-5}\ {\rm s}\times (M_L/ M_{\odot})$, $M_L=10^4\ {M_{\odot}}$ is large enough for the two images to be sufficiently isolated in time. 
As the modification effect is controlled by $\beta \sigma(\omega)$, we introduce a characteristic frequency
\begin{align}
    f_{\rm c}&=\frac{1}{2\pi}\sqrt{\beta \rho_{-2, -1}}\\
    &\sim 6.54\ {\rm Hz}\left(\frac{\xi_0}{0.86\ {\rm pc}}\right)^{1/2}
    \left(\frac{10^{4}M_{\odot}}{M_L}\right)^{1/2}
    \left(\frac{\rho_{-2, -1}}{4.16\times 10^{-33}\ {\rm m^{-2}}}\right)^{1/2}
    \,, 
\end{align}
determined by the condition $\beta\sigma(\omega)=1$. 
In high frequency region with $\beta \sigma(\omega) < 1$, the modification to the time delay rapidly decays.
Basically, for a smaller value of $\mu^2$
the characteristic frequency becomes lower. 
When we consider a mass term of the order of the current graviton mass constraint mentioned above, $f_c$ becomes significantly smaller than the observational band of ground-based GW detectors, resulting in no significant deviation from the GR predictions. 

\begin{figure}[htbp]
  \centering
  \begin{minipage}{0.45\textwidth}
    \includegraphics[width=\textwidth]{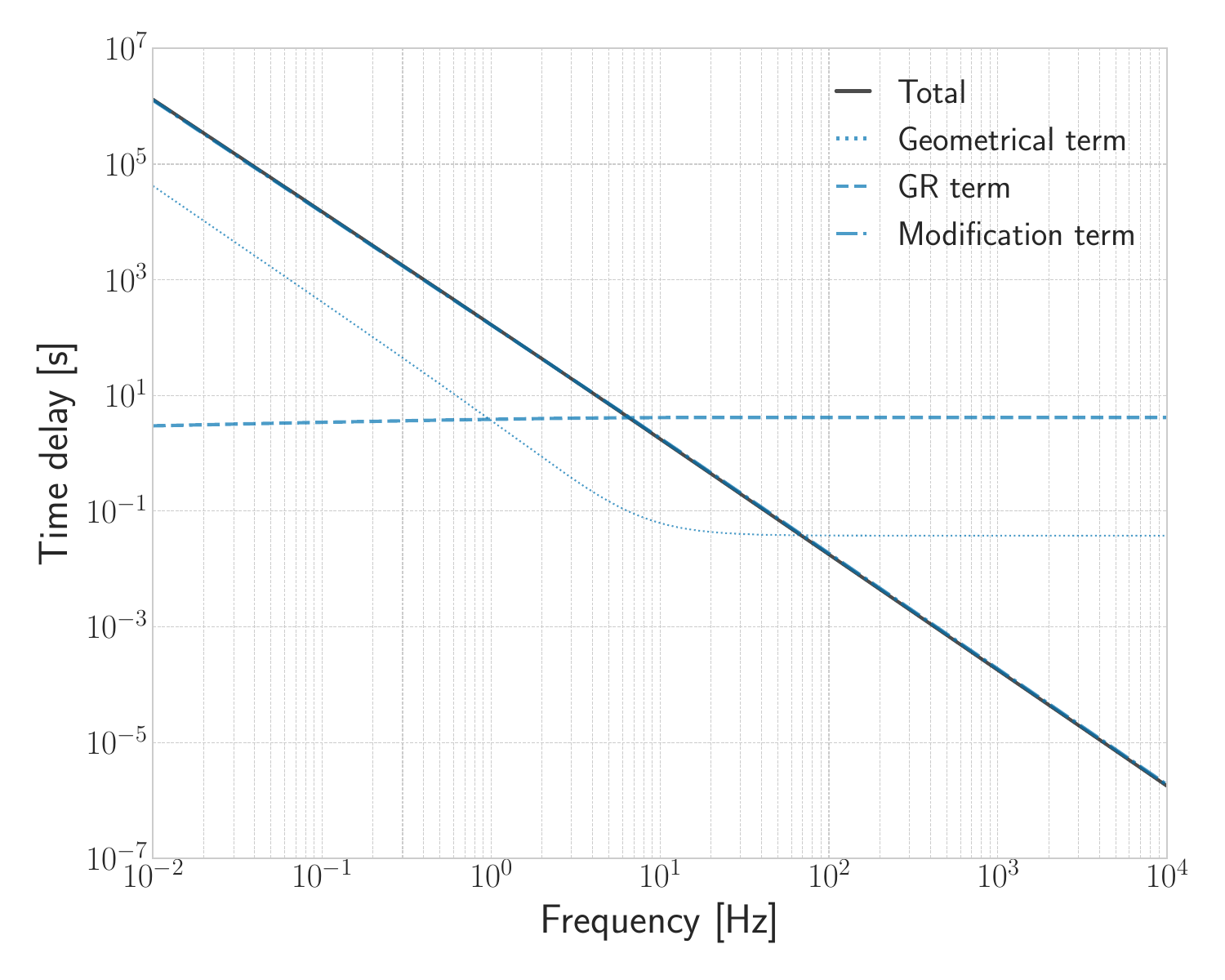}
  \end{minipage}
  \begin{minipage}{0.45\textwidth}
    \includegraphics[width=\textwidth]{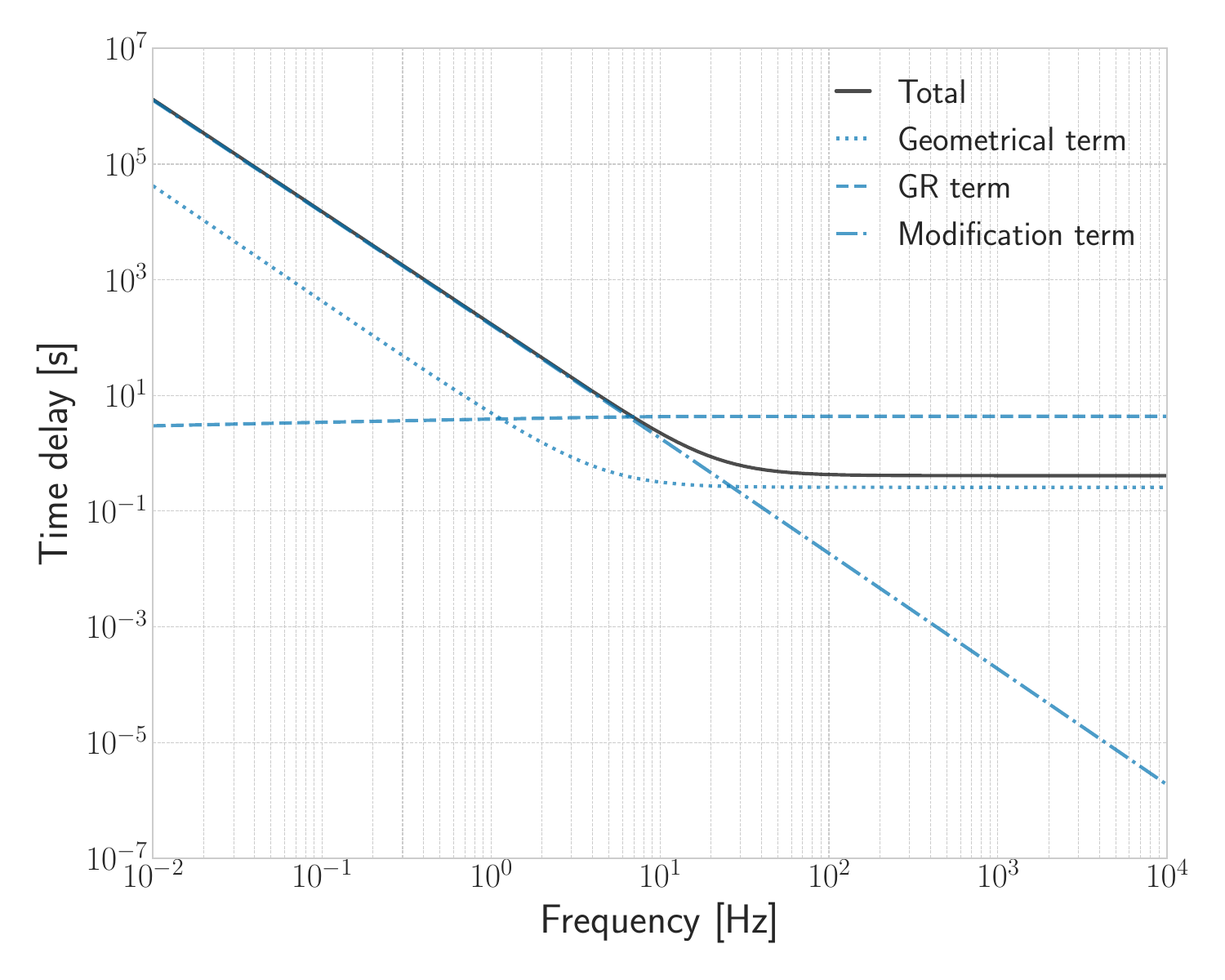}
  \end{minipage}
  \caption{Breakdown of the contributions to the total time delay for the plus image (left) and the minus image (right) with $\rho_{-2, -1}=-4.16\times 10^{-23}\ {\rm m^{-2}}$. The parameter setting is the same as in Fig.~\ref{fig:freq_vs_time_delay_spherical}.}
    \label{fig:freq_vs_time_delay_components_spherical}
\end{figure}

In low frequency region with $\beta \sigma(\omega) > 1$, time delay is modified, having frequency dependence $\propto \omega^{-2}$.
When $\rho_{-2, -1}$ is positive and the frequency is too low, the 
lens equation does not have a real solution, resulting in the absence of images at low frequencies. 
When $\rho_{-2, -1}$ is negative, there always exist solutions and hence images independent of frequency.
Figure~\ref{fig:freq_vs_time_delay_components_spherical} shows the contributions of the respective terms in Eq.~\eqref{eq:time_delay_spherical} to the time delay for $\rho_{-2, -1}= -4.16\times 10^{-23}$.
All terms, including the geometrical term $\propto (1/2) x_{\mp, S}^2$, the GR term $\propto -\log(x_{\pm, S})-(1/2)\log{(4D_L D_{LS}/\xi_0^2})$, and the modification term $\propto -\beta \sigma(\omega) [\log(x_{\pm, S})-(1/2)\log{(4D_L D_{LS}/\xi_0^2})]$ carry the same positive sign contribution.
The geometrical term also becomes frequency dependent because of the modification of the lens position as well as the modification term, which directly reflects the frequency dependence of $\beta\sigma(\omega)$. 
Although the GR term is also affected by the modification through the lens position, the effect is found to be minor in this case. 
In the case where $\rho_{-2, -1}$ is positive, images are lost at low-frequencies, but the behavior of the time delay is similar to the case when $\rho_{-2, -1}$ is negative. A difference is that, when $\rho_{-2, -1}$ is positive, the modification term contributes negatively, resulting in that the dip for the minus image appear due to the cancellation of the geometrical term and the potential terms.
Thus, the total time delays as depicted in the Fig.~\ref{fig:freq_vs_time_delay_spherical} are obtained.

From the eigenvalues of the Hessian matrix of the time delay, the magnification factors are calculated as  
\begin{align}
    \mu^{\rm mag}_{\pm, S}=\left(1-\frac{\left(1-\beta\sigma(\omega)\right)^2}{x_{\pm, S}^4}\right)^{-1}\,,
    \label{eq:magnification_factor_spherical}
\end{align}
in the geometrical optics regime. 

\begin{figure}
    \centering
    \includegraphics[width=3in]{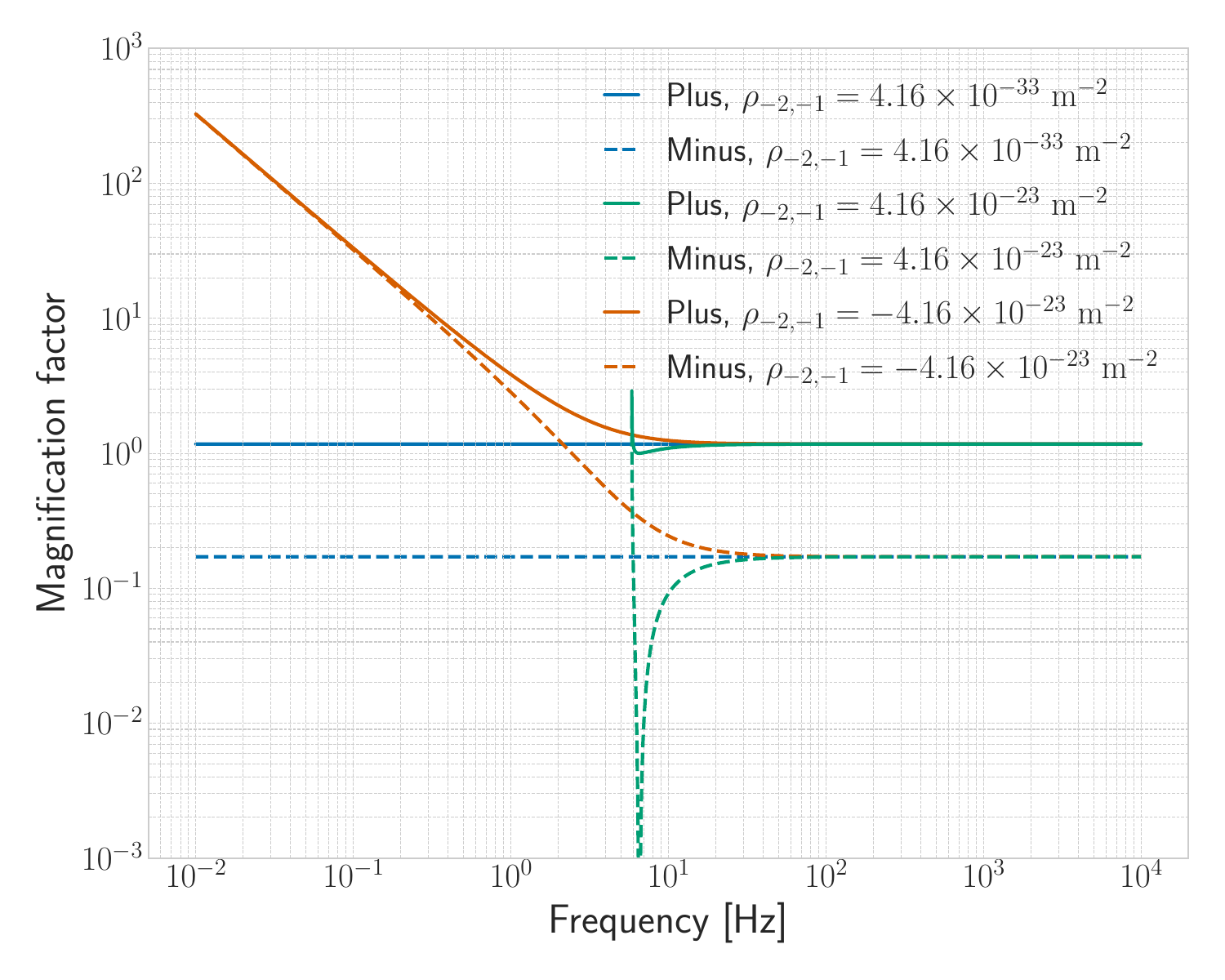}
    \caption{Same figure as Fig.~\ref{fig:freq_vs_time_delay_spherical} but for magnification factors.}
    \label{fig:freq_vs_magnification_factor_spherical}
\end{figure}

Figure~\ref{fig:freq_vs_magnification_factor_spherical} illustrates the frequency dependence of the magnification factor for three cases as before. 
Here as well, below the characteristic frequency, the effects of the modification become apparent, and converge to the values predicted by GR on the higher frequency side. 
When $\rho_{-2, -1}$ is positive, images disappear at low frequencies, but when $\rho_{-2, -1}$ is negative, magnification occurs on the low-frequency side.
The fact that the modifications for both plus and minus images match on the lower frequency side indicates that the modification through $\beta\sigma(\omega)$ in Eq.~\eqref{eq:magnification_factor_spherical} dominates over the modification through the image position.
Therefore, depending on the values of parameters included in $\beta$ and $\sigma(\omega)$, the local modification in GW propagation equation can change not only the time delay but also the magnification in frequency dependent manners maintaining the unlensed waveform consistent with the GR waveform, which cannot be achieved by cosmological uniform modifications.

\begin{figure}[htbp]
  \centering
  \begin{minipage}{0.48\columnwidth}
    \includegraphics[width=\columnwidth]{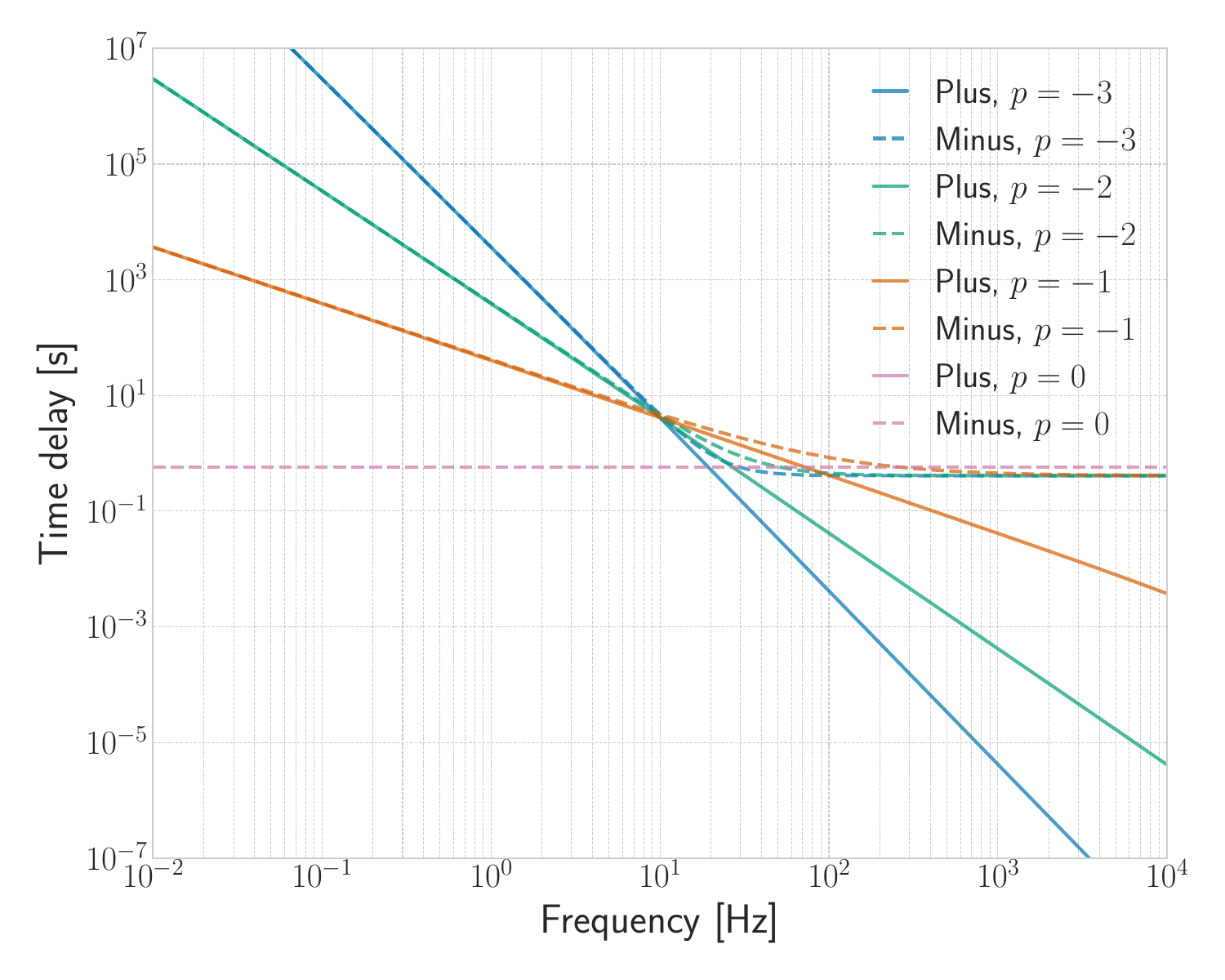}
  \end{minipage}
  \begin{minipage}{0.48\columnwidth}
    \includegraphics[width=\columnwidth]{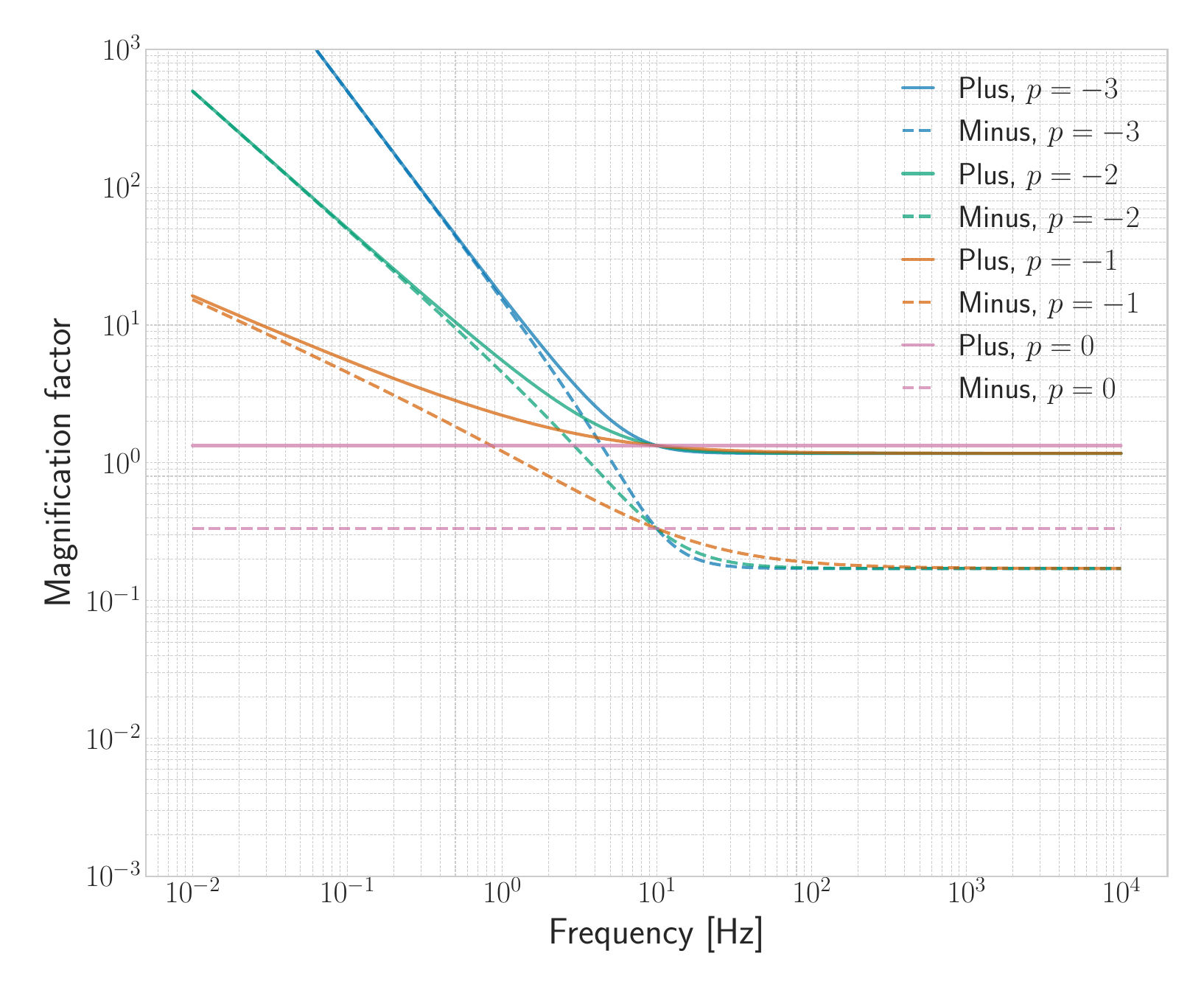}
  \end{minipage}
  \caption{Time delays (left) and magnification factors (right) for different negative powers where the absolute values of the negative $\rho_{p, -1}$ so that the characteristic frequency gives $10\ {\rm Hz}$. The fiducial values of the lensing parameters are $z_S = 0.5$, $z_L=0.1$, $M_L=10^4\ {M_{\odot}}$, and $y=1.0$.}
  \label{fig:negative_power_comparison}
\end{figure}

\begin{figure}[htbp]
  \centering
  \begin{minipage}{0.48\columnwidth}
    \includegraphics[width=\columnwidth]{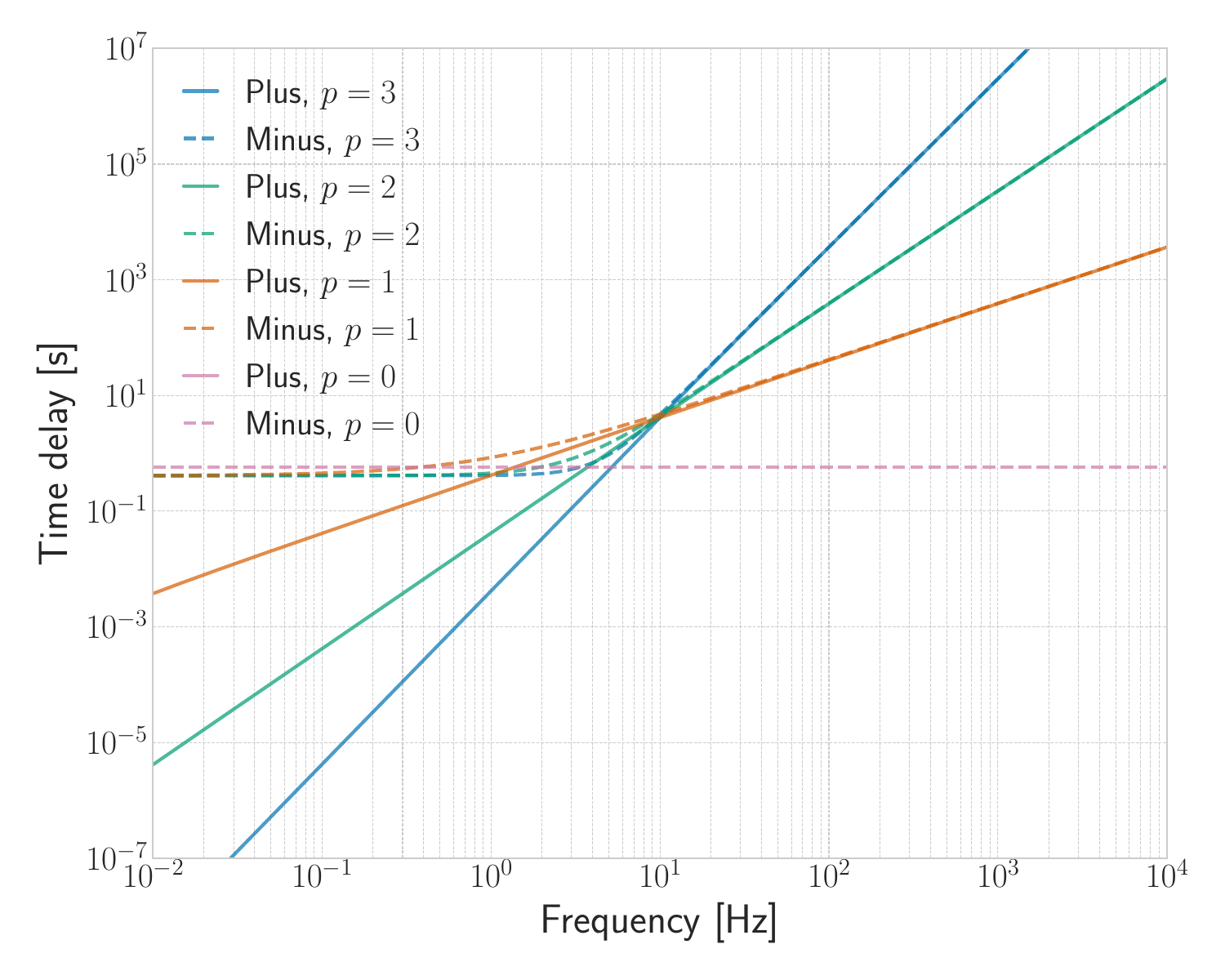}
  \end{minipage}
  \begin{minipage}{0.48\columnwidth}
    \includegraphics[width=\columnwidth]{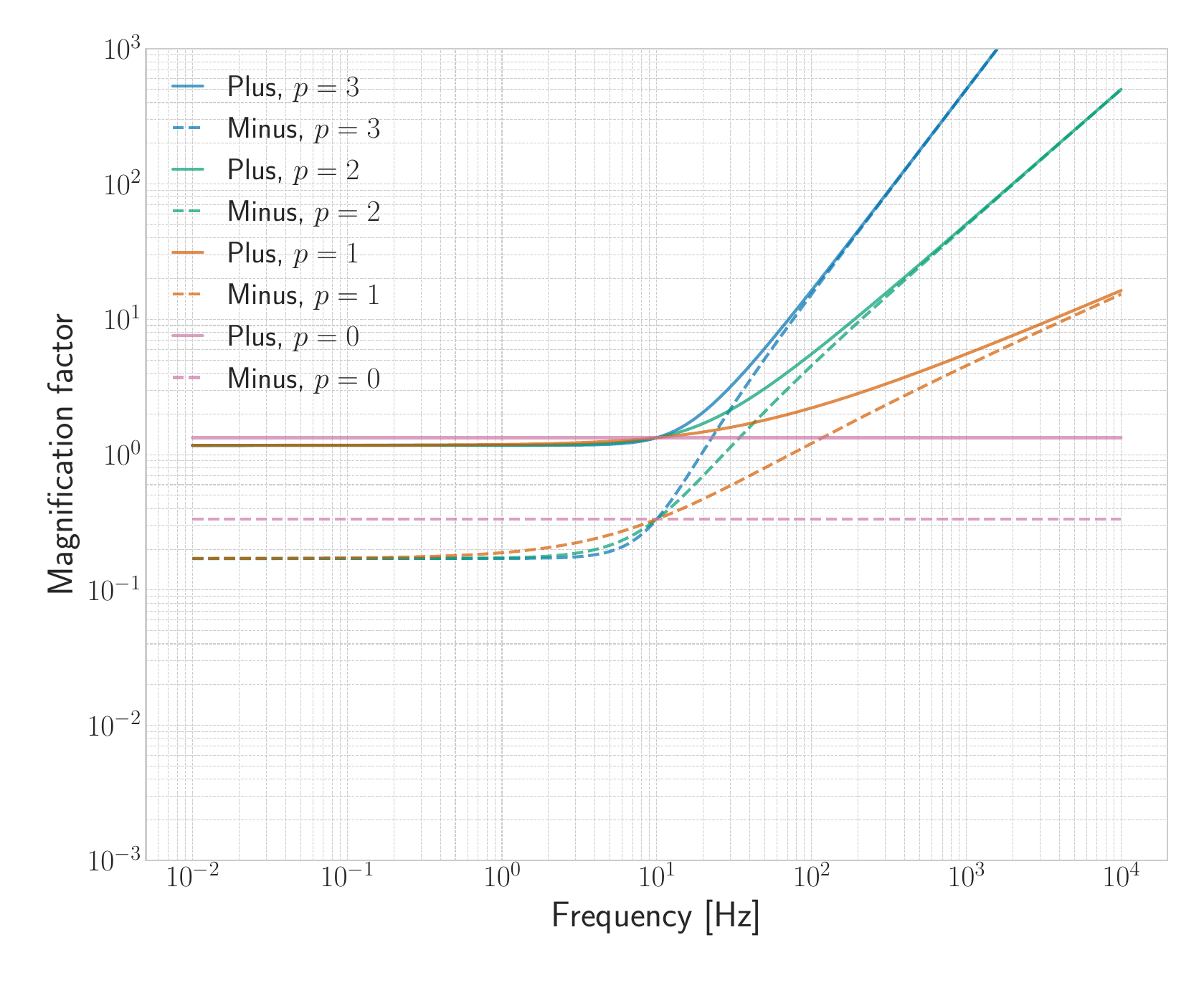}
  \end{minipage}
  \caption{Time delays (left) and magnification factors (right) for different positive powers where the absolute values of the negative $\rho_{p, -1}$ so that the characteristic frequency gives $10\ {\rm Hz}$. The fiducial values of the lensing parameters are $z_S = 0.5$, $z_L=0.1$, $M_L=10^4\ {M_{\odot}}$, and $y=1.0$.}
  \label{fig:positive_power_comparison}
\end{figure}

Next, we consider the scenario where the frequency dependence of the local modification is varied, $\sigma(\omega) = \rho_{p, -1} \omega^{p}$.
We show the cases with negative $\rho_{p, -1}$, for which the images exist for all frequencies. 
Figure~\ref{fig:negative_power_comparison} shows the time delays and the magnification factors for negative powers, while Fig.~\ref{fig:positive_power_comparison} compares those for positive powers.
The fiducial parameters are $z_S = 0.5$, $z_L=0.1$, $M_L=10^4\ {M_{\odot}}$, and $y=1.0$ again. 
The absolute values of the modification $\rho_{p, -1}$ are adjusted so that the characteristic frequency $f_{\rm c}=(1/2\pi)(\beta \rho_{p, -1})^{1/p}$ gives $10\ {\rm Hz}$.
The effects induced by $\sigma(\omega) = \rho_{p, -1} \omega^{p}$ are similar to those caused by $\mu$.
Below these characteristic frequencies when $p$ is negative or above these characteristic frequencies when $p$ is positive, the time delay and magnification factor exhibit extra frequency dependence $\propto f^{p}$.
Although $p=0$ or constant $c_{\phi}$ itself does not introduce frequency dependence to the time delay or magnification factor, speed deviations around the lens object increase the time delay and magnification factor across all frequency ranges.
Recall that the modifications due to $c_\phi^2$ and $\mu^2$ terms correspond to 
$p=-0$ and $p=-2$, respectively. 

Therefore, local corrections to the propagation of GWs enable polarization-selective and frequency dependent magnification. The local magnifications keep the unlensed waveform consistent with the GR waveform, which cannot be achieved with cosmological modifications. 
Both the time delay and magnification factor reflect the frequency dependence of the modifications.
When $M^2$ is positive, the images disappear in the frequency bands where the modification is large. On the other hand, when $M^2$ is negative, GWs could be frequency-dependently amplified.
From the characteristic frequency dependence of the amplitude and phase for lensed GWs, it may be possible to deduce the type of local modifications. 
However, signals significantly magnified by such modifications are expected to be deformed in a frequency-dependent manner, raising questions about their detectability within the current search pipelines that primarily use GR templates. 
The detectability of these effects will be discussed in the following section.

\section{Detectability}
\label{sec:detectability}
In this section, we discuss the detectability of lensed GWs with spherically symmetric modifications to the the propagation equation. 
First, we assess the signal to noise ratio to investigate how the signal can be magnified due to the local modifications. 
Second, we evaluate the mismatch between the modified waveforms and the GR templates.

\subsection{Signal to noise ratio}
The matched filter signal-to-noise ratio (SNR) for a signal $s(t) = h(t) + n(t)$, composed of a GW strain $h(t)$ and noise $n(t)$, against a template $h_{T}$ is defined by
\begin{align}
    \rho = \frac{(s|h_T)}{\sqrt{(h_T|h_T)}}\simeq\frac{(h|h_T)}{\sqrt{(h_T|h_T)}}\,,
    \label{eq:matched_filter_SNR}
\end{align}
where we define the inner product for the Fourier components as
\begin{align}
    (A|B) = 4 {\rm Re} \left[ \int^{\infty}_{0}\frac{\tilde{A}^*(f)\cdot\tilde{B}(f)}{S_n(f)}df \right]\,,
\end{align}
with the single-sided power spectral density $S_{n}(f)$. 
The optimal SNR $\rho_{\rm opt}$ is obtained when the template matches the GW strain $h_T\propto h$,
\begin{align}
    \rho_{\rm opt}=\sqrt{(h|h)}\,.
\end{align}
A loss of SNR by $x\%$ roughly implies that only $(1-x/100)^3$ of total events can be detected due to the reduction in the detector horizon.
We shall investigate the SNR of the lensed GWs with the local modifications. 
As we are interested in the comparison with the GWs without modifications, we use the relative SNR to the lensed GWs in GR, $\rho_{\rm opt}/\rho_{\rm opt, GR}$ where $\rho_{\rm opt}$ is the optimal SNR for the lensed GWs with modifications and $\rho_{\rm opt, GR}$ is that calculated in GR without modifications. 

\begin{figure}[htbp]
      \centering
  \begin{minipage}{0.48\columnwidth}
    \includegraphics[width=\columnwidth]{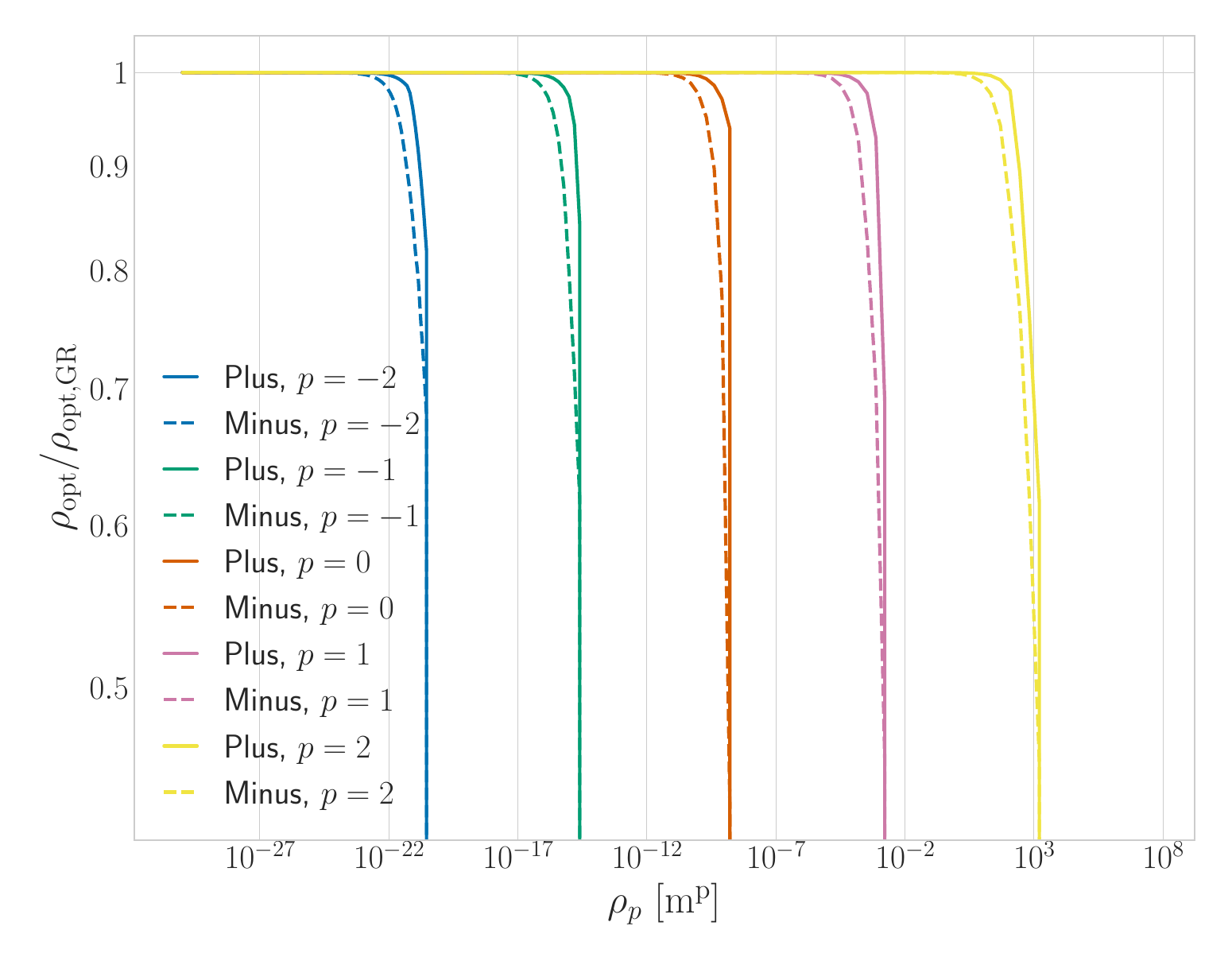}
  \end{minipage}
  \begin{minipage}{0.48\columnwidth}
    \includegraphics[width=\columnwidth]{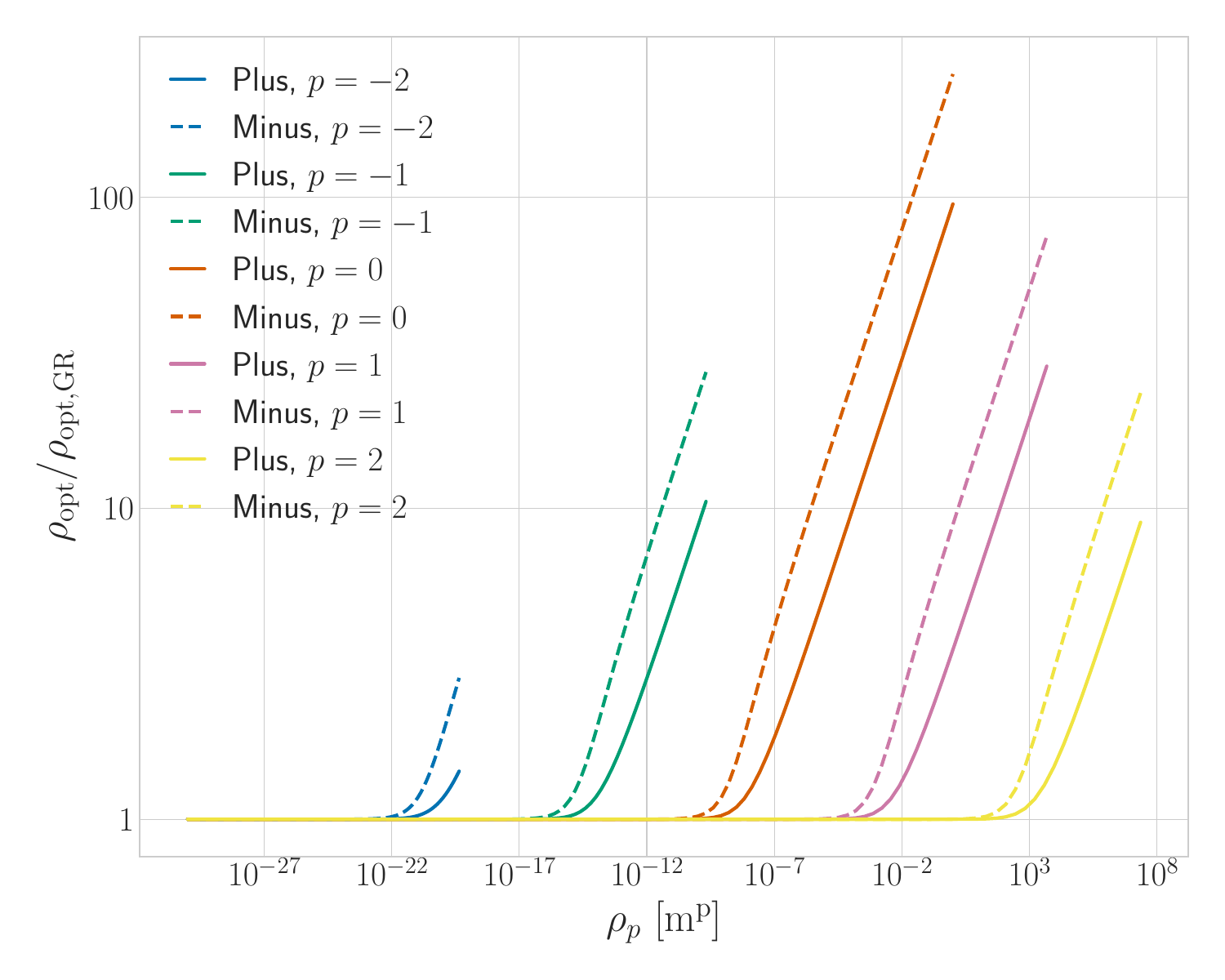}
  \end{minipage}
  \caption{Relative signal-to-noise ratio of the image with corrections of $\rho_{p, -1}\omega^{p}$ to that in GR for positive $\rho_{p, -1}$ (left) and negative $\rho_{p, -1}$ (right).
  The fiducial parameters of the lensing system are given by $z_S = 0.5$, $z_L=0.1$, $M_L=10^4\ {M_{\odot}}$, and $y=1.0$. The binary parameters are set to the chirp mass $\mathcal{M}=26.1\ M_{\odot}$, the symmetric mass ratio $\eta=0.25$, the effective spin $\chi_{\rm eff}=0$, and the inclination angle $\iota=0.0\ {\rm rad}$.}
  \label{fig:snr_comparison}
\end{figure}

Figure~\ref{fig:snr_comparison} shows the relative SNR for the lensed image with the corrections of the form $\rho_{p, -1} \omega^p$.
We are plotting the relative SNR within the approximation where the correction $|\rho_{p, -1} \omega^p|$ is less than unity, ensuring that the modification remains small compared to the contributions from GR.
The lensing parameters are set to the fiducial values $z_S = 0.5$, $z_L=0.1$, $M_L=10^4\ {M_{\odot}}$, and $y=1.0$.
The binary parameters are given by the chirp mass $\mathcal{M}=26.1\ M_{\odot}$, the symmetric mass ratio $\eta=0.25$, the effective spin $\chi_{\rm eff}=0$, and the inclination angle $\iota=0.0\ {\rm rad}$.
We have marginalized over the antenna pattern response in GW signal and used the design sensitivity of advanced LIGO.

When $\rho_{p, -1}$ is positive, it is observed that the SNR decreases due to the loss of images in certain frequency bands. 
As the value of $\rho_{p, -1}$ is varied, the characteristic frequency crosses the detector's observational band, altering the extent of SNR reduction. 
On the other hand, when $\rho_{p, -1}$ is negative, the SNR increases due to frequency-dependent amplification caused by the local modifications.
The relative magnification is larger for the minus image, and the increase of SNR starts at the magnitude when the characteristic frequency $f_{c} = (1/2\pi)(\beta \rho_{p, -1})^{1/p}$ crosses the detector's observational band. 
As the absolute value of $p$ increases, the maximum magnification ratio decreases, within the range where the modification $|\rho_{p, -1} \omega^p|$ is less than unity.
For instance, a greater maximum magnification is expected for local deviations in the speed of GW propagation ($p=0$) than for local modifications of mass ($p=-2$).
While magnification due to the ordinary gravitational potential is equally provided to all GW polarizations, the modifications beyond GR could depend on the polarization. 
Therefore, local modifications may allow for polarization-selective magnification.

\subsection{Waveform mismatch}
We have investigated how modifications in GW propagation can deform the waveform of lensed GWs, and found that sometimes the SNR is amplified. 
However, significant alterations to the waveform can hinder efficient event detection using GR templates. 
Here, we assess the waveform mismatch between the lensed waveform with modifications and the original waveform in GR for a single detector.

Combining the matched filter SNR and the optimal SNR, we can define the faithfulness $F$, which is a measure indicating agreement of the GW waveform against the template,
\begin{align}
    F = \frac{\rho}{\rho_{\rm opt}}=\frac{(h|h_T)}{\sqrt{(h|h) (h_T|h_T)}}\,,
\end{align}
and the mismatch $\mathscr{M}$, which is a measure indicating disagreement of the two waveforms,
\begin{align}
    \mathscr{M} = 1-F\,.
\end{align}
According to the Lindblom criterion~\cite{Lindblom:2008cm, Chatziioannou:2017tdw} requiring that the true parameters are less than $1\sigma$ away from the best fit ones, two waveforms are considered indistinguishable if 
\begin{align}
\mathscr{M}<\frac{D}{2 \rho_{\rm opt}^2}\,,
\end{align}
where $D$ is the number of parameters involved in the waveforms. 
Now, we have $D=6$ intrinsic parameters: $(\mathcal{M}, \eta, \chi_{\rm eff}, M_L, y, \rho_p)$.
Hence, for the detection of signals deformed by local modifications using GR templates with $\rho_{\rm opt}=8$, $50$, and $100$, a mismatch $\mathscr{M}$ of less than or equal to $5\%$, $0.1\%$, and $0.01\%$, respectively.

\begin{figure}[htbp]
      \centering
  \begin{minipage}{0.48\columnwidth}
    \includegraphics[width=\columnwidth]{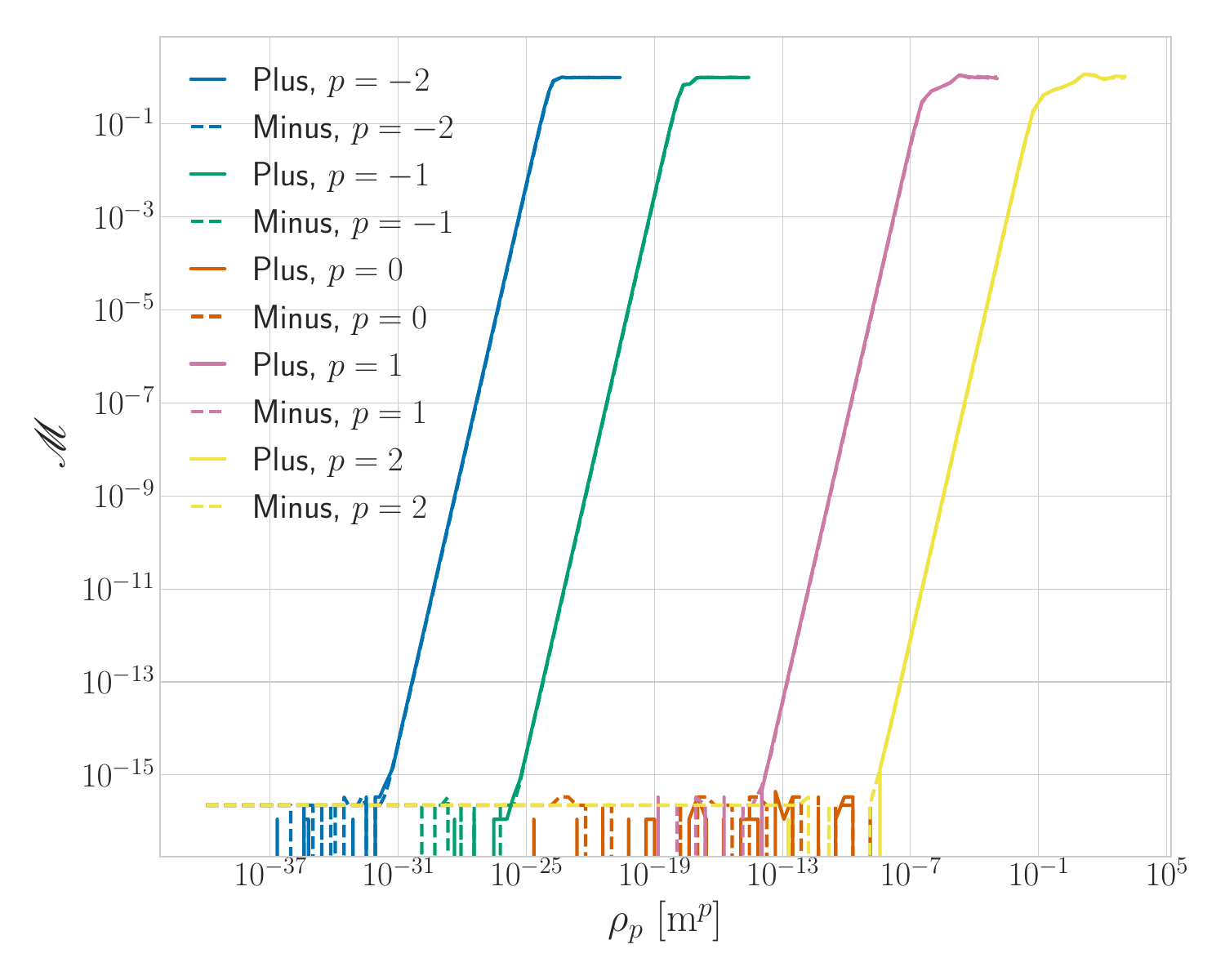}
  \end{minipage}
  \begin{minipage}{0.48\columnwidth}
    \includegraphics[width=\columnwidth]{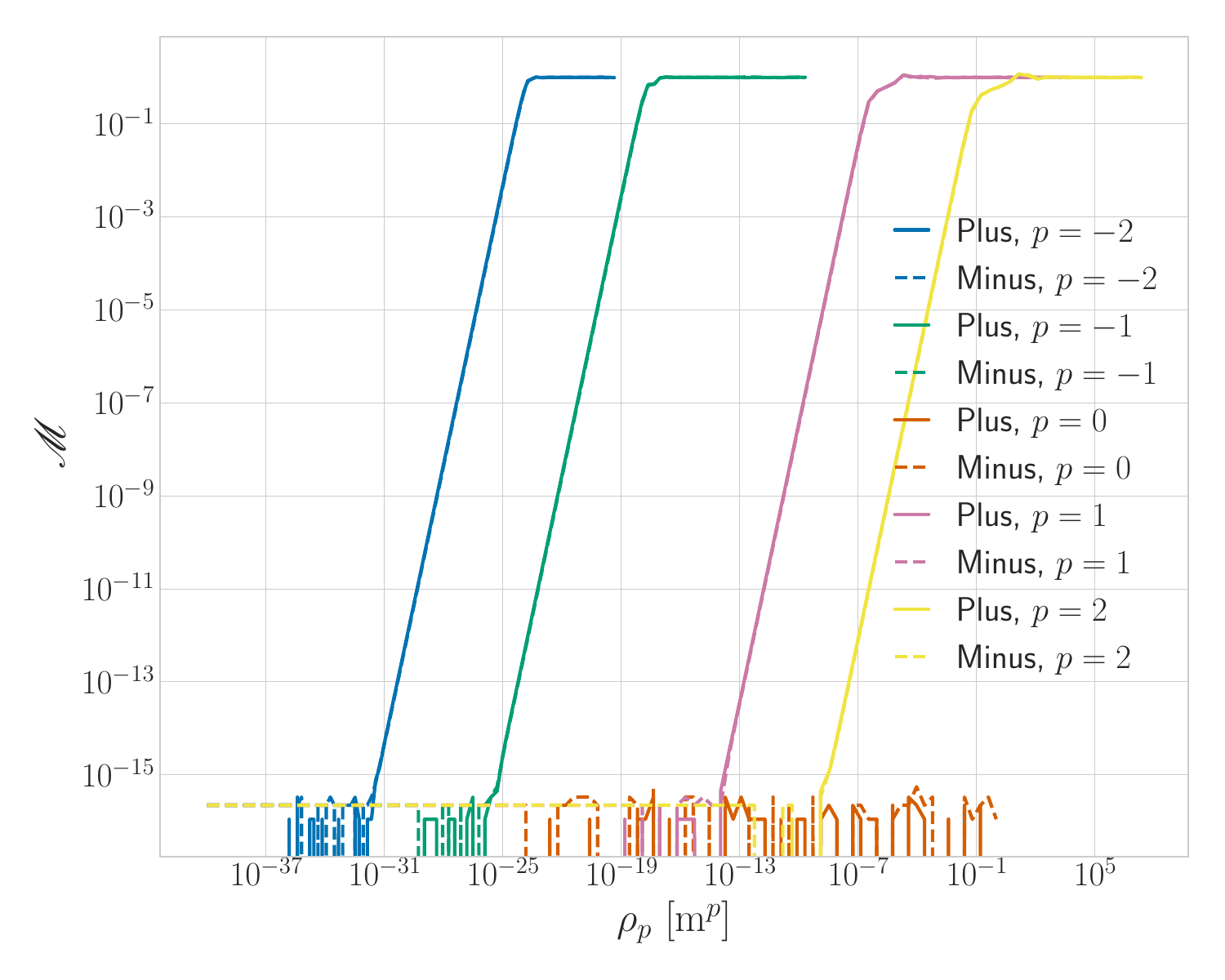}
  \end{minipage}
    \label{fig:rho_vs_mismatch_any_power}
  \caption{Waveform mismatch between the lensed waveform with modifications of $\rho_{p, -1} \omega^p$ and the waveform in GR for positive $\rho_{p, -1}$ (left) and negative $\rho_{p, -1}$ (right). The fiducial parameters of the lensing system are given by $z_S = 0.5$, $z_L=0.1$, $M_L=10^4\ {M_{\odot}}$, and $y=1.0$. The binary parameters are set to the chirp mass $\mathcal{M}=26.1\ M_{\odot}$, the symmetric mass ratio $\eta=0.25$, the effective spin $\chi_{\rm eff}=0$, and the inclination angle $\iota=0.0\ {\rm rad}$.}
\end{figure}

Figure~\ref{fig:rho_vs_mismatch_any_power} illustrates the mismatch between the lensed waveform with $\rho_{p, -1} \omega^p$ and the GR waveform.
The parameter setting is the same as in Fig.~\ref{fig:snr_comparison}.
The lensing parameters are set to the fiducial values.  
The binary parameters are given by the chirp mass $\mathcal{M}=26.1\ M_{\odot}$, the symmetric mass ratio $\eta=0.25$, the effective spin $\chi_{\rm eff}=0$, and the inclination angle $\iota=0.0\ {\rm rad}$.
As the lensed waveform with a non-zero $c_{\phi}$ $(p=0)$ is identical to that predicted by GR aside from constant magnifications, the mismatch is zero, irrespective of the parameter value.
For the case of $p\neq0$, the waveform undergoes significant deformation, accompanied by either an increase or a decrease in SNR due to local modifications. 
This leads to a larger value of mismatch in both the positive and negative $\rho_{p, -1}$ cases.
Particularly in the parameter regions where a significant magnification is expected, the value of the mismatch exceeds the Lindblom criterion, making detection with GR templates challenging. 
It should be noted that for tensor modes in regions without extra amplification, detection using GR templates remains feasible. 
Thus, events detected with GR templates can be used to explore a wide range of parameter regions by means of the parameterized waveform model.

The converse of the Lindblom criterion does not hold generally.
Namely, because of the degeneracies and the biases arising from correlations between parameters, the value of mismatch exceeding the threshold does not necessarily guarantee undeniable detection of the modifications. 

\section{Discussion}
\label{sec:discussion}
Until now, we have been assuming a form of $\propto 1/r$, for local corrections in the propagation equation. 
If we consider a variety of models of modifications, different shapes can be obtained. 
For instance, the effective field theory of gravity predicts the correction $\propto 1/r^6$ to the propagation speed of GWs~\cite{deRham:2020ejn}. 
Gravitons may gain an exponentially damping mass localized around black holes~\cite{Zhang:2017jze}.
Thus, depending on the local modification models and the local structure of the lens, diverse patterns of time delay and magnification could be induced.
Therefore, here we consider different power-law profiles of the local modifications. 

Let us consider the local modification profile $\propto r^n$.
In this case, the correction to the deflection potential $\psi_{M}({\bm x})$ is determined by the function $S_n(\bm{x}, \ell)$ specified by Eq.~\eqref{eq:S_n}.
The image position is given by the lens equation,
\begin{align}
    y = x -\frac{1}{x} \mp \beta \sigma(\omega) \frac{\partial P_n(|x|)}{\partial |x|}\,.
\end{align}
The equation with $-$ sign determines the image positions on the source side, while that with $+$ sign determines the image positions on the other side.
Given the conditions $x \ll D_L/\xi_0$ and $x \ll D_{LS}/\xi_0$, the function $ \partial P_n/\partial x$ behaves approximately as $\propto |x|^n$. 
Consequently, under these conditions, the lens equation becomes an $n$-th order equation in terms of $x$ for a given $y$,
\begin{align}
    y-x + \frac{1}{x} = \mp c_n \beta \sigma(\omega)|x|^{n}\,,
\end{align}
where $c_n$ are constant coefficients, {\it e.g.},  $c_{-1}=-1,\ c_{-2}=-\pi/2,\ c_{-3}=-2,\ c_{-4}=-3\pi/4,\ c_{-5}=-8/3,\ c_{-6}=-15\pi/16, \cdots$. 
In the following, similarly to what has been discussed so far, $x>0$ will represent solutions on the source side, while $x<0$ will denote solutions on the other side.

\begin{figure}[htbp]
  \centering
  \begin{minipage}{0.48\columnwidth}
    \includegraphics[width=\columnwidth]{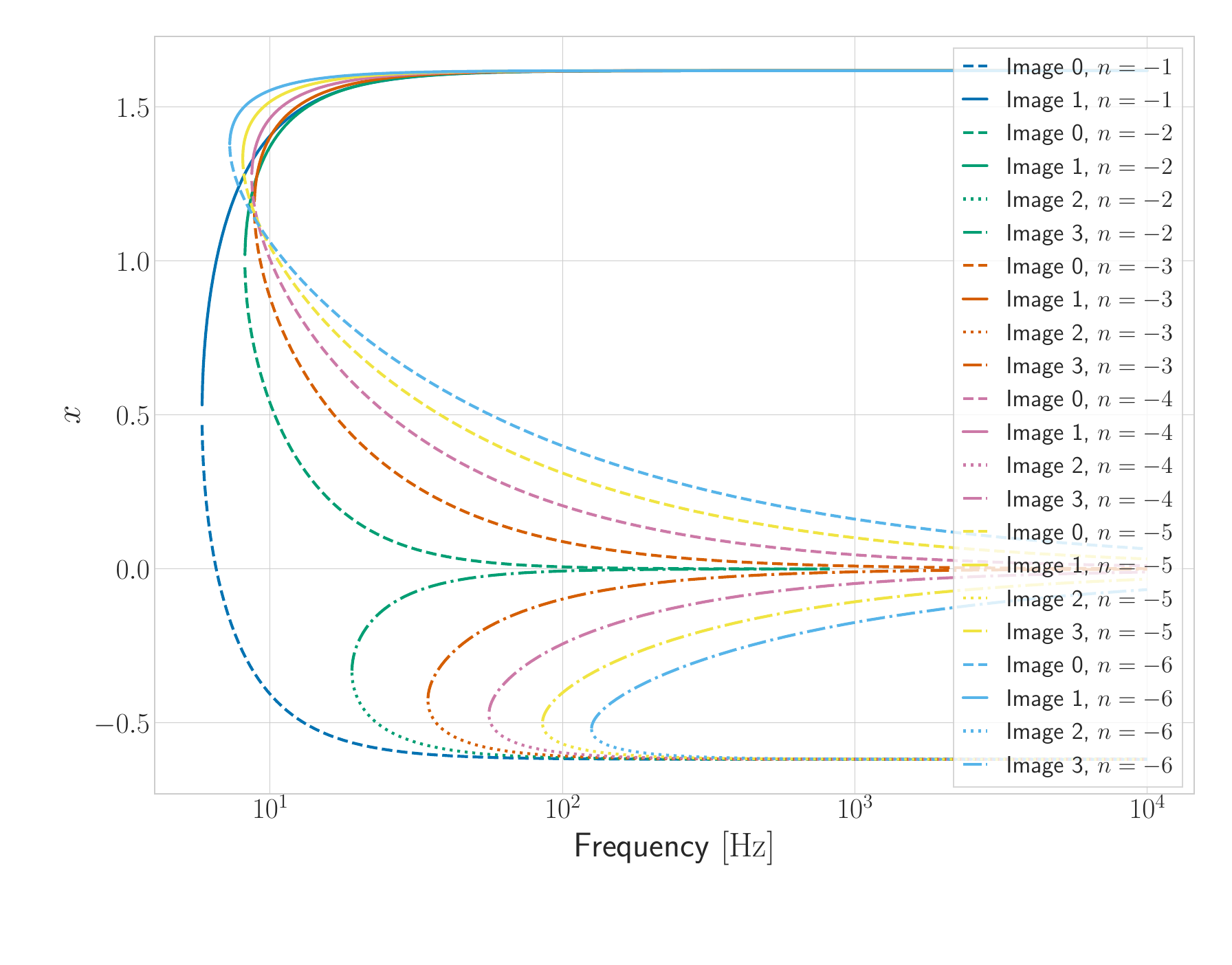}
  \end{minipage}
  \begin{minipage}{0.48\columnwidth}
    \includegraphics[width=\columnwidth]{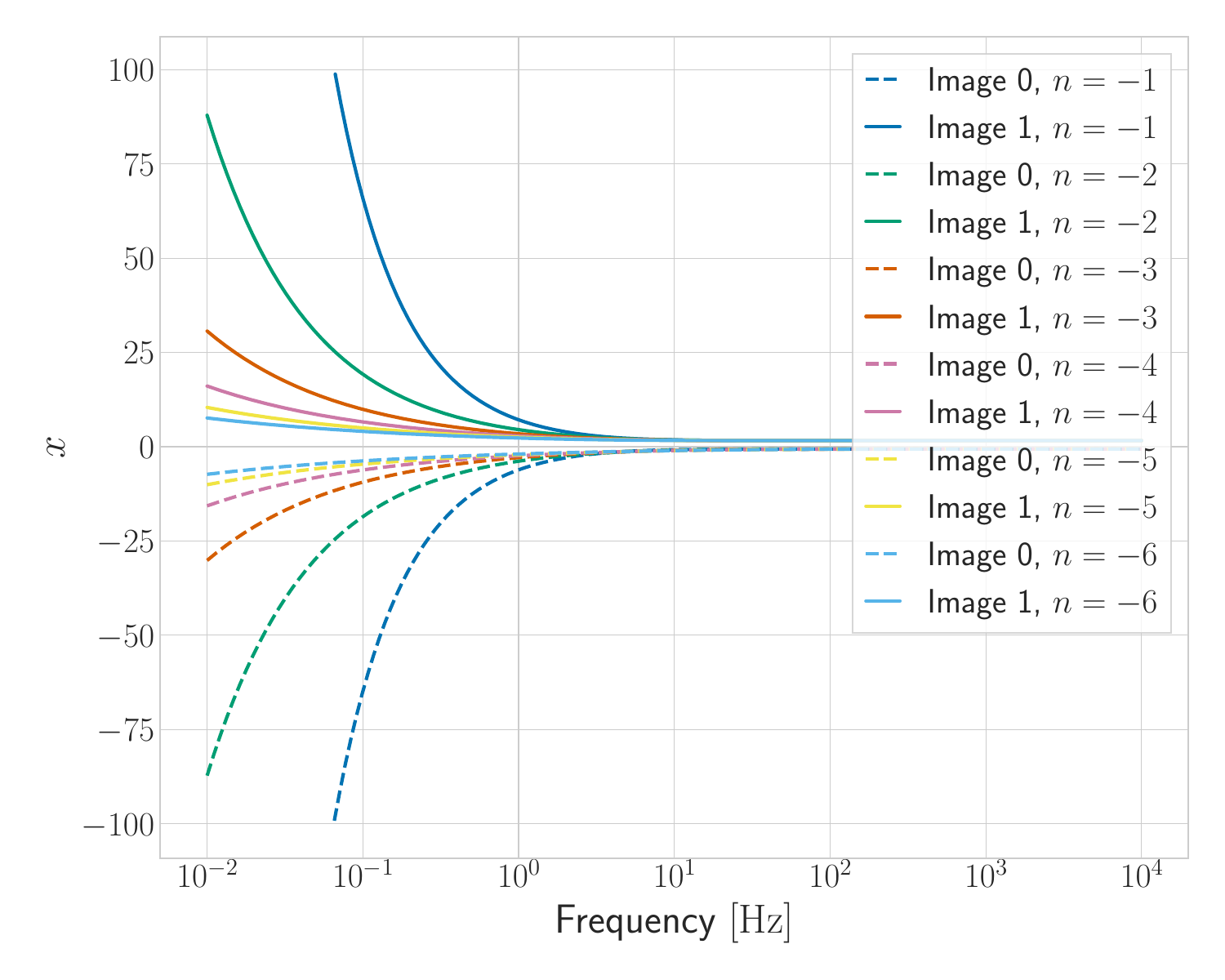}
  \end{minipage}
  \caption{Frequency dependence of the image position $x$ for $\rho_{-2, n}=4.16\times 10^{-23}\ {\rm m^{-2}}$ (left) and $\rho_{-2, n}=-4.16\times 10^{-23}\ {\rm m^{-2}}$ (right).
  The lensing parameters are given by the fiducial values $z_S = 0.5$, $z_L=0.1$, $M_L=10^4\ {M_{\odot}}$, and $y=1.0$.}
  \label{fig:freq_vs_x}
\end{figure}

Figures~\ref{fig:freq_vs_x} illustrates the frequency dependence of image positions for different $n$ for $ \rho_{-2, n}=\pm 4.16\times 10^{-23}\ {\rm m^{-2}}$ as we assumed in Figs.\ref{fig:freq_vs_time_delay_spherical}-\ref{fig:freq_vs_magnification_factor_spherical}.
The lensing parameters are set to the fiducial values. 
When $\rho_{-2, n}$ is positive, solutions exist only above the characteristic frequency, and no images are produced on the low-frequency side. 
The number of images on the high-frequency side changes depending on $n$ as the frequency increases. 
For $n=-1$, two images are always obtained, but for the other $n$, up to four images can be obtained. 
When $\rho_{-2, n}$ is negative, two images are always obtained.

\begin{figure}[htbp]
  \centering
  \begin{minipage}{0.48\columnwidth}
    \includegraphics[width=\columnwidth]{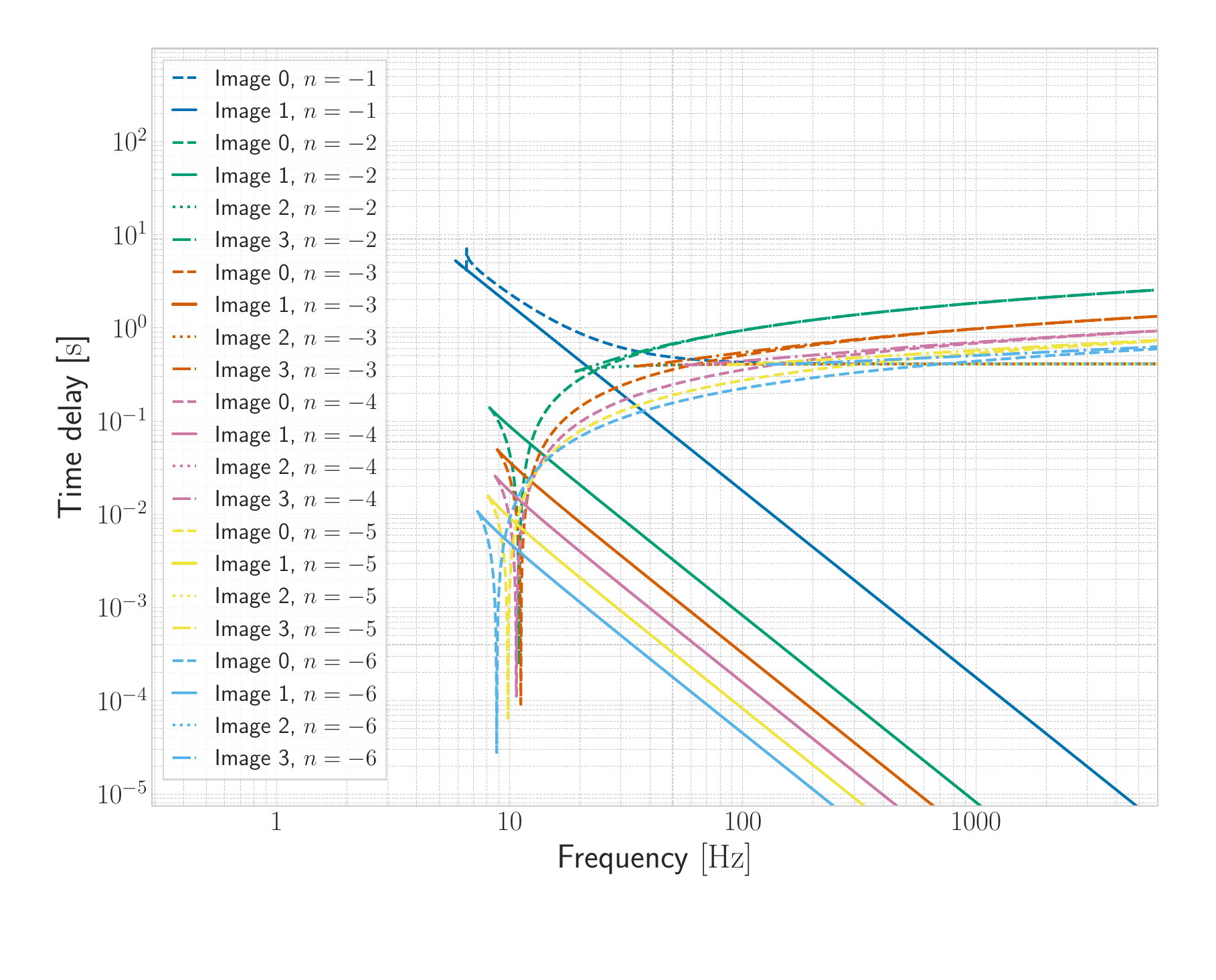}
  \end{minipage}
  \begin{minipage}{0.48\columnwidth}
    \includegraphics[width=\columnwidth]{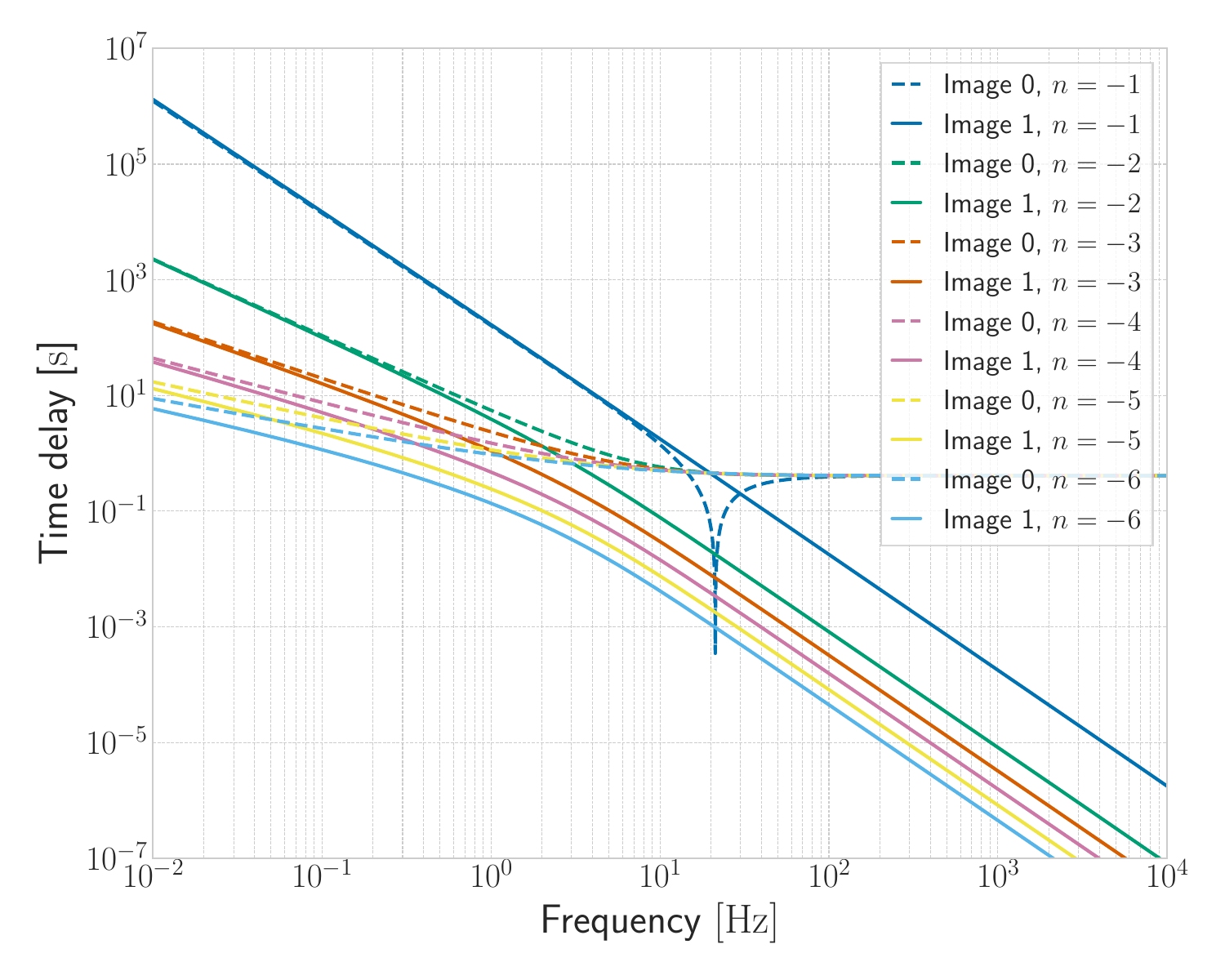}
  \end{minipage}
  \caption{Frequency dependence of the time delay for each image with $\rho_{-2, n}=4.16\times 10^{-23}\ {\rm m^{-2}}$ (left) and $\rho_{-2, n}=-4.16\times 10^{-23}\ {\rm m^{-2}}$ (right).
  The lensing parameters are given by the fiducial values $z_S = 0.5$, $z_L=0.1$, $M_L=10^4\ {M_{\odot}}$, and $y=1.0$.}
  \label{fig:freq_vs_time_delay_spherical_any_n}
\end{figure}

The time delay for the images can be calculated using the effective deflection potential ~\eqref{eq:effective_deflection_potential_power_law}. 
Figure~\ref{fig:freq_vs_time_delay_spherical_any_n} shows the frequency dependence of the time delays for the images varying the power law index $n$ with the same value of $\rho_{-2, n}$ as in Figs.\ref{fig:freq_vs_time_delay_spherical}-\ref{fig:freq_vs_magnification_factor_spherical}. 
Here, we use the time delay at the high frequency for the 1-st image as the baseline.
For $n=-1$ the 0-th image corresponds to the minus image and the 1-st image corresponds to the plus image in Sec.~\ref{sec:time_delay_and_magnification_factor}.
Note that, below the characteristic frequency for negative $\rho_{-2,n}$, the modification term becomes dominant, while above that, the frequency dependence becomes very weak. 
For positive $\rho_{-2,n}$, the frequency dependence on the high frequency side still remains (image 0 and 3). 
This is because the images are very close to the center $(x=0)$ where the lensing potential is very steep, and hence the time delay is very sensitive to the small change in the image positions. 
When $\rho_{-2, n}$ is positive, the time delays only have values in the regions where the image exists, reflecting the appearance and disappearance of the images.
When $n = -1$, the time delays  are as shown in Fig.~\ref{fig:freq_vs_time_delay_spherical}. 
For $n \leq -2$, the time delay for the image 0 decreases towards the characteristic frequency from high-frequency side, with a dip appearing due to the cancellation of different terms. 
For the image 1, the time delay increases toward the characteristic frequency, and decreases monotonically as $|n|$ increases. 
The time delay for the image 2 asymptotically approaches the time delay for the image 0 on the high-frequency side. 
The time delays for the image 3 remains the same regardless of the value of $n$.
When $\rho_{-2, n}$ is negative, the time delays are similar to that for $n = -1$.
In this case again, as $|n|$ increases, the frequency-dependence becomes weaker, 
and hence the observational effects are expected to be more suppressed.

On calculating the determinant of the Hessian matrix of the time delay, the magnification factor is derived as
\begin{align}
    \mu_{\rm mag}(\bm{x}_j)= \left| \frac{x_j^4}{(-1+|x_j|^2-\beta \sigma(\omega) |x_j| P'_n)(1+|x_j|^2-\beta \sigma(\omega) |x_j|^2 P''_n)} \right|\,,
\end{align}
where the prime indicates the differentiation with respect to $|x|$.

\begin{figure}[htbp]
  \centering
  \begin{minipage}{0.48\columnwidth}
    \includegraphics[width=\columnwidth]{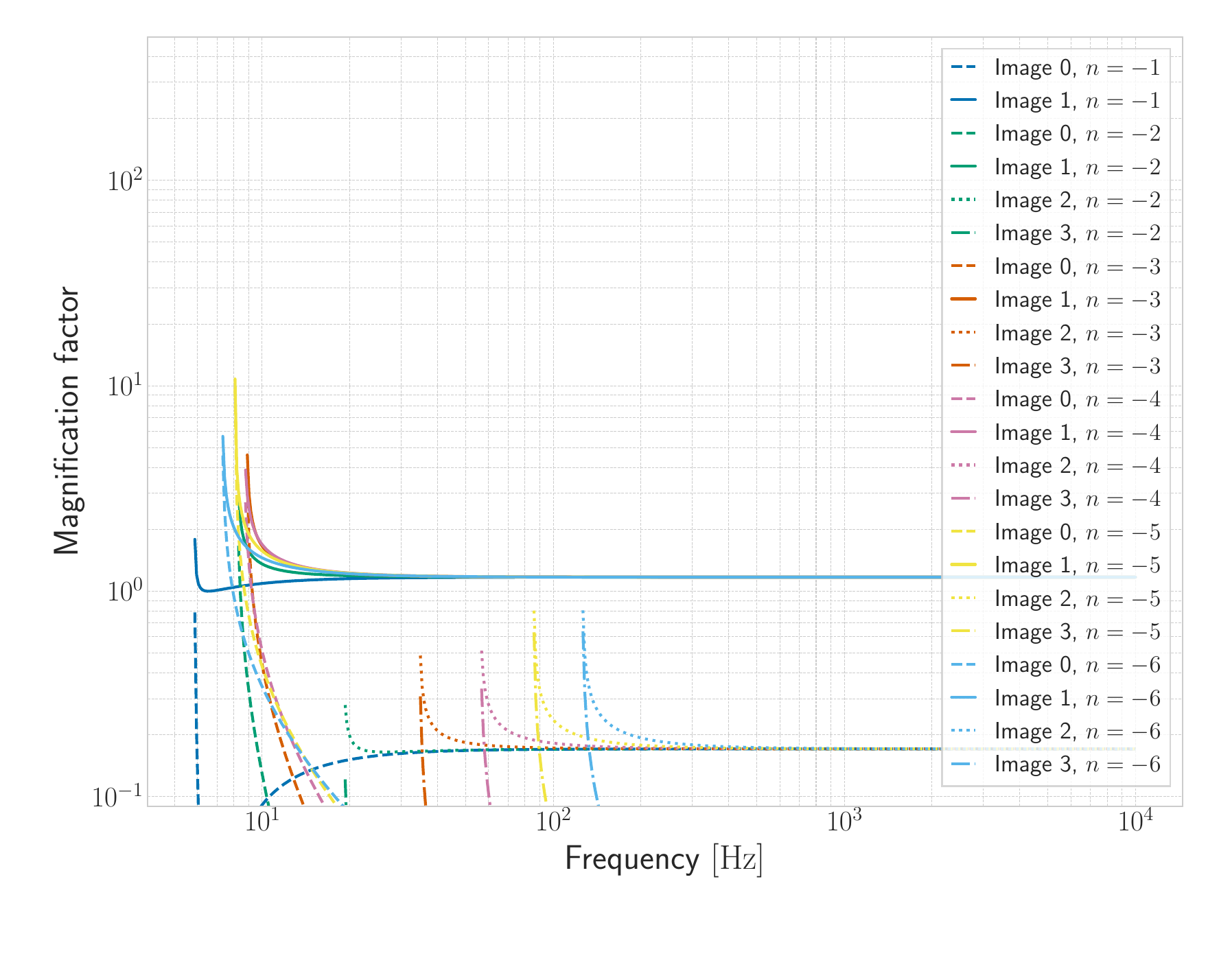}
  \end{minipage}
  \begin{minipage}{0.48\columnwidth}
    \includegraphics[width=\columnwidth]{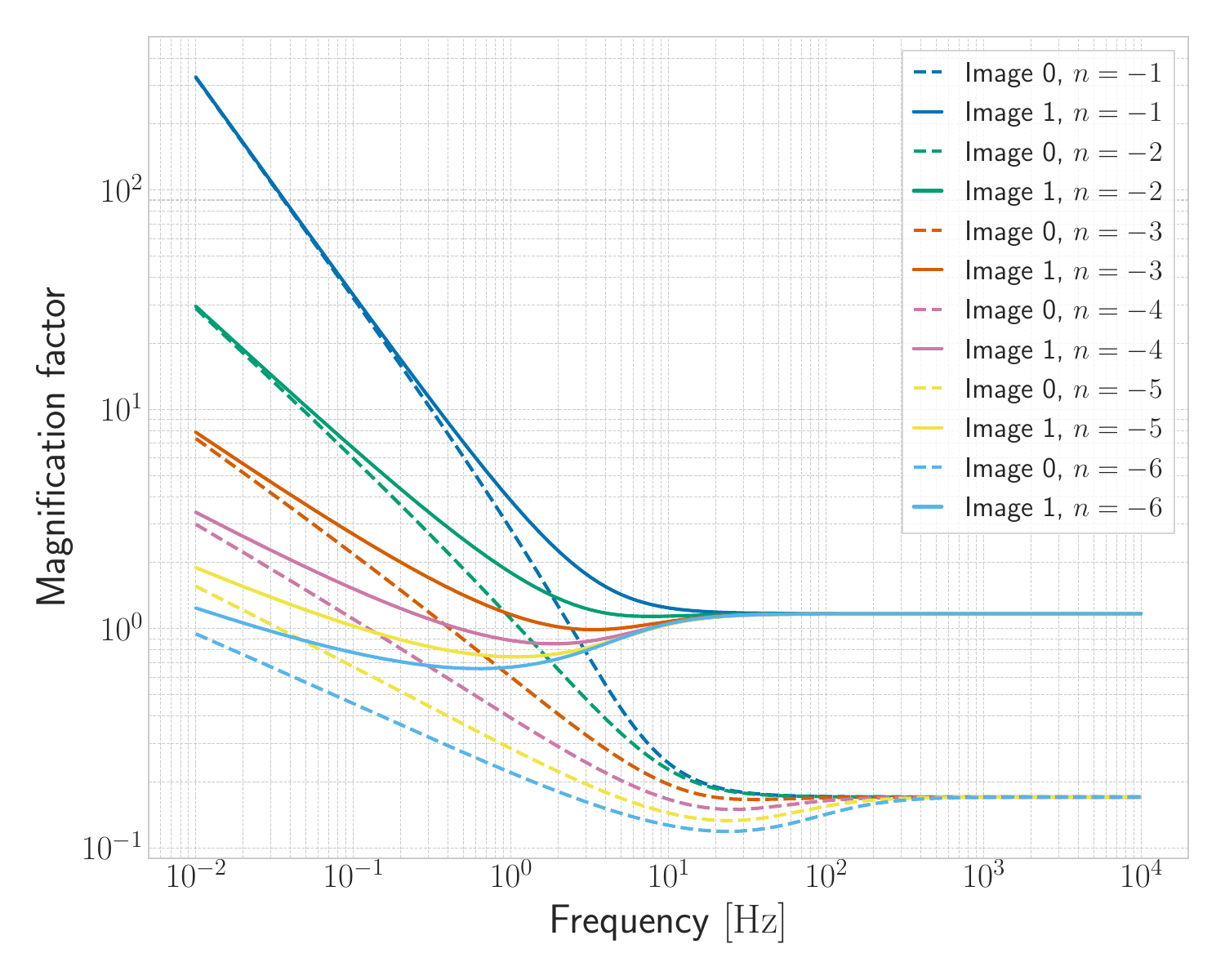}
  \end{minipage}
  \caption{Frequency dependence of the magnification factor for each image with $\rho_{-2, n}=4.16\times 10^{-23}\ {\rm m^{-2}}$ (left) and $\rho_{-2, n}=-4.16\times 10^{-23}\ {\rm m^{-2}}$ (right).
  The lens parameters are given by the fiducial values $z_S = 0.5$, $z_L=0.1$, $M_L=10^4\ {M_{\odot}}$, and $y=1.0$.}
  \label{fig:freq_vs_magnification_factor_spherical_any_n}
\end{figure}

Figure~\ref{fig:freq_vs_magnification_factor_spherical_any_n} shows the magnification factor of each image for various $n$ in frequency domain 
with the same value of $\rho_{-2, n}$ as in Figs.\ref{fig:freq_vs_time_delay_spherical}-\ref{fig:freq_vs_magnification_factor_spherical}.  
For positive $\rho_{-2, n}$, except for $n=-1$, although up to four images can be obtained, 
for two of these images (images 0 and 3), the peaks of the magnification factors are significantly localized in the frequency space, which are comparable to images 1 and 2 only within a limited frequency domains. 
Consequently, observing images 0 and 3 would be challenging, although we mentioned their weak frequency dependence of the time delay in the above.
For negative $\rho_{-2, n}$, low-frequency components are amplified as well as the case with $n=-1$.
As expected, the frequency dependence weakens as $|n|$ increases.

In summary, the number and nature of images change depending on the signature of $\rho_{-2,n}$.
When $\rho_{-2, n}$ is positive, the images disappear on the low-frequency side, and up to four distinct images with different frequency dependence appear on the high-frequency side. However, for two of these (image 0 and image 3), the magnification factors are localized in the frequency domain, meaning that they are comparable to images 1 and 2 only within a limited frequency range. 
Hence, observing images 0 and 3 would be challenging.
Therefore, detecting images 1 and 2, which lack only their low-frequency components, becomes observationally crucial for probing positive corrections, regardless of the value of $n$.
On the other hand, when $\rho_{-2, n}$ is negative, the observation of two images with amplified low-frequency components is expected. 
In this case, as $|n|$ increases, the frequency dependence is suppressed.
We have illustrated the image positions, time delay, and magnification factor for integer values of $n$, but the behavior is similar and changes continuously for non-integer values of $n$.
Here, we present the case for $p = -2$, but a similar discussion can be made for negative values of $p$. 
For positive values of $p$, the argument can be made by swapping the low-frequency side with the high-frequency side, and when $p = 0$, no frequency dependence is introduced.
Therefore, by analyzing the frequency dependence of the images of lensed GWs, it may be possible to distinguish the signs of $\rho_{p, n}$ and the values of $n$.

\section{Conclusion}
\label{sec:conclusion}
We investigated the effects of modifications to the GW propagation equation on the waveform of gravitationally lensed GWs, under the geometrical optics approximation. 
The corrections can be categorized into two types: uniform cosmological modifications that are continuously at work along all the way from the generation to the detection of GWs, and local corrections that are restricted to a small region around the lensing objects. 
Adopting a phenomenological approach that parameterizes the GW propagation equation without assuming a specific theoretical model, we primarily explored the effects of these modifications in the realm of strong lensing, where multiple images are separated in time. 
If we consider the cases in which two images interfere, we need to take into account complex beat patterns. 
In this paper, we set the lens parameters using the image separation condition at the optimal frequency for the ground-based GW telescopes, and evaluated the SNR and mismatch for individual images. 
However, in cases where there is significant amplification on the low-frequency side, frequency-dependent interference may occur, necessitating the investigation of beat patterns for actual analysis.
In addition, here we restricted our consideration basically to the propagation without mode mixing. Models that can have a mode mixing would produce some unique signatures.
It would be an intriguing avenue for further investigation
to explore how such effects could influence the discussions presented in this paper.

For cosmological modifications, the overall factor of the time delay due to gravitational lensing is affected through the modified phase velocity, but this effect is negligibly small compared to the effects that accumulate through all the way of the cosmological long-distance propagation, which exist already in the unlensed waveform.
The cosmological modifications can also contribute to the effective deflection potential.
However, for tensor modes, if there is a significant magnification due to this effect, some signatures should have been already visible in the tests of GW propagation, and hence existing observations would be enough to exclude this possibility. 
On the other hand, for non-tensorial modes, the deformation of the unlensed waveform would become too severe to be detected, before the additional amplification becomes significant.
Thus, the advantage of probing cosmological modifications with lensing systems lies not in the unique feature in the modified waveforms, but in gaining a longer propagation distance thanks to the signal magnification caused by the lensing.

In contrast, local corrections can modify the time delays and hence the image positions and magnification factors in a frequency-dependent manner through corrections to the deflection potential.
We pointed out the presence of polarization-selective and frequency-dependent effects: image disappearance for positive effective masses, and signal amplification for negative effective masses.
Assuming 
local corrections with a simple radial profile such that gives the same $r-$dependence of the deflection potential as the ordinary point mass lense, we evaluate the magnification of the signal-to-noise ratio for lensed GW signals.  
Lensed images can disappear below the characteristic frequency, or the signal can be amplified by an order of magnitude, 
but we found that the waveform in such cases are significantly deformed from GR ones by evaluating the mismatch between the modified and the un-modified waveforms. 

Frequency-dependent amplification based on local modifications to the GW propagation equation motivates not only tests of GW using tensor modes of GWs, but also search for non-tensorial modes. 
Generally, metric theories of gravity allow up to four scalar and vector modes in addition to the two tensor modes, and searching for non-tensorial modes can be a compelling direction to probe deviations from GR. 
However, existing experiments indicate that some mechanism of magnification is necessary to make extra polarizations detectable~\cite{Takeda:2023mhl}.
Gravitational lensing could serve as one such mechanism. 
While the ordinary gravitational lensing effects contribute equally to all modes of GWs, the modified propagation can provide different amplification to each mode, depending on the structure of the background fields around massive objects. 
Local modifications in propagation could lead to polarization-selective amplification, but result in modifications to the waveform at the same time. 
Therefore, when searching for anomalous polarizations, it is necessary to consider a search strategy that takes into account modifications to the waveform.

\section*{Acknowledgements}
We would like to thank Katsuki Aoki, Atsuhi Nishizawa, and Ryuichi Takahashi for useful comments.
H.T. is supported by Japan Society for the Promotion of Science (JSPS) KAKENHI Grant Nos. JP21J01383 and JP22K14037 and by the Hakubi project at Kyoto University.
T.T. is supported by JSPS KAKENHI Grant Nos. JP24H00963, JP24H01809, JP23H00110 and JP20K03928.

\appendix
\section{Gravitational-wave polarization and detector signal}
\label{sec:polarization_and_signal}
The GW signals obtained from GW detectors are observable quantities, and hence they should be diffeomorphism-invariant and gauge-invariant. 
Therefore, the GW signal $h(t)$ can be expressed as a contraction of the GW tensor $S_{ij}$ and the detector tensor $d_{ij}$,
\begin{align}
    h(t)=d^{jk}S_{jk}\,.
\end{align}
In metric theories of gravity that satisfy Einstein's equivalence principle, the metric perturbation and the stress-energy tensor of matter universally couple,
\begin{align}
    S_{\rm int} = \int d^{4}x \frac{\sqrt{-g}}{2}h_{\mu\nu}T^{\mu\nu}\,.
\end{align}
Under this coupling, the motion of test masses constituting the detector is described by the geodesic equation. 
In the case of interferometric GW telescopes, the observable quantity is the phase difference of laser lights combined at the beam splitter. 
Solving the geodesic deviation equation for the test masses, we obtain the detector tensor
\begin{align}
    d^{jk}=\frac{1}{2}\left( \hat{x}_{d}^{j}\hat{x}_{d}^{k}-\hat{y}_{d}^{j}\hat{y}_{d}^{k}  \right)
\end{align}
where $\hat{x}_{d}$ and $\hat{y}_{d}$ are unit vectors along the arms of the detector,
and the GW tensor which is related to the Riemann tensor as~\cite{Maggiore:2007ulw, Poisson2014}
\begin{align}
    R_{0j0k}=-\frac{1}{2}\partial_{0}^2 S_{jk}\,.
\end{align}
where $\partial_{0}$ stands for the derivative in terms of the time coordinate.
In the synchronous gauge, as the component of the Riemann tensor is reduced to $R_{0j0k}=(1/2)\ddot{h}_{ij}$,
the GW signal modulo a constant is simply given by 
\begin{align}
    h(t)=d^{jk}h_{jk}\,.
    \label{eq:GW_signal}
\end{align}
 After substituting Eq.~\eqref{eq:propagating_modes} into Eq.~\eqref{eq:GW_signal} and defining the GW polarization $h_A$,
\begin{align}
    h_A := \sum_{I} \phi_{I} \xi^{I}_{A}\,,
\end{align}
and the antenna patterns functions 
\begin{align}
    F^{A}:=d^{ij}e_{ij}^{A}\,,
\end{align}
we immediately rewrite the GW signal as the linear combination of the GW polarization modes multiplied by the antenna pattern functions~\cite{Tobar:2009sf, Takeda:2018uai, Takeda:2019gwk, Nishizawa:2009bf},
\begin{align}
    h(t)=\sum_{A}h_{A}F^{A}\,.
\end{align}
Explicit forms of the antenna pattern functions are provided in~\cite{Nishizawa:2009bf, Poisson2014}, and please refer to~\cite{Ezquiaga:2020gdt} for differences in conventions of angular parameters.

\bibliography{bib}

\end{document}